# Equivariant cohomology and gauged bosonic $\sigma$-models

José M Figueroa-O'Farrill[†] and Sonia Stanciu[‡§]

*Department of Physics, Queen Mary and Westfield College*
*Mile End Road, London E1 4NS, UK*

ABSTRACT

We re-examine the problem of gauging the Wess–Zumino term of a $d$-dimensional bosonic $\sigma$-model. We phrase this problem in terms of the equivariant cohomology of the target space and this allows for the homological analysis of the obstruction. As a check, we recover the obstructions of Hull and Spence and also a generalization of the topological terms found by Hull, Roček and de Wit. When the symmetry group is compact, we use topological tools to derive vanishing theorems which guarantee the absence of obstructions for low dimension ($d \leq 4$) but for a variety of target manifolds. For example, any compact semisimple Lie group can be gauged in a three-dimensional $\sigma$-model with simply connected target space; or any compact semisimple Lie group having no cubic casimirs can be gauged in a four-dimensional $\sigma$-model with a target space a Lie group with at most one $U(1)$ factor. When the symmetry group is semisimple but not necessarily compact, we argue in favor of the persistence of these vanishing theorems by making use of (conjectural) equivariant minimal models (in the sense of Sullivan). By way of persuasion, we construct by hand a few such equivariant minimal models, which may be of independent interest. We illustrate our results with two examples: $d = 1$ with a symplectic target space, and $d = 2$ with target space a Lie group admitting a bi-invariant metric. An alternative homological interpretation of the obstruction is obtained by a closer study of the Noether method. This method displays the obstruction as a class in BRST cohomology at ghost number 1. We comment on the relationship with consistent anomalies.

[†] e-mail: J.M.Figueroa@qmw.ac.uk

[‡] e-mail: S.Stanciu@qmw.ac.uk

[§] On leave of absence from Physikalishem Institut der Universität Bonn.
Address after October 1994: ICTP, Trieste, Italy.

## §1 Introduction

The nonlinear $\sigma$-model has proven to be one of the most fascinating examples of field theories. Its physical applications range from the description of the low-energy interactions of hadrons in four dimensions (see **[1]** and references therein), to the construction of string backgrounds, namely conformally invariant two-dimensional field theories. Moreover it plays a fundamental role in two-dimensional conformal field theory by providing a lagrangian description of known (conjecturally all) rational conformal field theories. At a more formal level, the nonlinear $\sigma$-model plays a privileged role in mathematical physics. It bridges easily between physics and geometry: usually a problem in one discipline can be translated into a problem in the other via a $\sigma$-model description. Indeed, it has been used to model geodesic motion on a Riemannian manifold, Morse theory, mirror symmetry, index theorems,... and it is more and more becoming part of the bag of tricks of physics-aware mathematicians.

Our interest in this paper is the study of symmetries of (nonlinear) bosonic $\sigma$-models with Wess–Zumino (WZ) term. The original WZ term **[2]** is a nonlinear interaction term for scalar fields taking values in a Lie group $G$—that is, for maps from spacetime to $G$—and its purpose was to cancel the chiral anomaly in a fermionic theory with $(G \times G)$-symmetry. In two dimensions, the analogous object gives rise to the celebrated Wess–Zumino-Witten (WZW) model **[3]**. In two-dimensions and for a special value of the coupling, there is an explosion of symmetry and in fact, the WZW model is invariant under the loop group of $G \times G$, roughly speaking. Moreover the WZW model does not just mimic certain aspects of a fermionic theory, but is fully equivalent to one, namely to the Thirring model.

Whenever one has a symmetry it is natural to gauge it. Usually there is no obstruction to this: one introduces gauge fields and substitutes derivatives for covariant derivatives, but as we will see later on, gauging the WZ term in a $\sigma$-model presents us with obstructions that are to be overcome. It is familiar that for the two-dimensional WZW model only "anomaly-free" subgroups of the full $G \times G$ symmetry group can be gauged. For a general two-dimensional $\sigma$-model with a WZ term, the obstructions to gauging were analyzed in **[4]** and independently in **[5]**. In **[6]** Hull and Spence tackled the general problem and wrote down a series of obstructions that had to be overcome in order to gauge a $d$-dimensional $\sigma$-model with WZ term.

The motivation for our present work stems from the work of Hull and Spence and grew out of a desire to identify the precise cohomology theory responsible for the obstructions in **[6]** with the aim of deriving a priori vanishing theorems that would guarantee the absence of obstruction in certain cases of interest. In this paper we do exactly this. We show that the obstructions in [**6**] correspond to a single class in a subcomplex of the equivariant cohomology theory of the target space. This allows us to guarantee the absence of obstructions to gauging a $d$-dimensional $\sigma$-model with WZ term for large classes of target manifolds and for dimensions $d \leq 4$.

This paper is organized as follows. In Section 2 we introduce the problem. We define what we mean by gauging a bosonic $\sigma$-model and we show that the principal term can be gauged simply by minimal coupling. We then show that this does not work for the Wess–Zumino term; hence presenting us with obstructions which we will later interpret in terms of equivariant cohomology. We end the section with some topological heuristics that support the role played by equivariant cohomology. This last subsection can be skipped at a first reading—its purpose being mainly to motivate the machinery developed in the next section.

In Section 3 we introduce the Weil algebra and the Weil homomorphism and exhibit its role in the de Rham model for equivariant cohomology. In this framework we see that equivariance is the same as gauge-invariance and that (topological) gauge-invariant terms for the $\sigma$-model action are nothing but equivariant cocycles. This means that gauging a Wess–Zumino term is finding an equivariant extension to a given invariant form.

This is exploited in Section 4 to study the obstructions cohomologically. We define a subcomplex of the equivariant forms on the target manifold in which the obstruction lies. In fact this new complex fits in a short exact sequence of complexes giving rise to a long exact sequence in cohomology. The obstruction to gauging the Wess–Zumino term turns out to be precisely the image under the connecting homomorphism of the class of the Wess–Zumino term in the invariant cohomology of the target space. We conclude the section with a cohomological look at the Noether method. This allows us to exhibit the obstruction as a class in BRST cohomology at ghost number one. We comment on the relation with the homological description of consistent anomalies.

In Section 5 we use the Cartan model for equivariant cohomology, to expand the universal obstruction found in Section 4 and thus recover the obstructions found by Hull and Spence in [**6**]. Examining the ambiguities in the gauged action once the obstructions have been overcome, we associate new topological terms to the $\sigma$-model to certain equivariant cocycles. These topological terms include and generalize the topological terms found by Hull, Roček and de Wit in [**7**].

In Section 6 we illustrate the nature of these obstructions with two examples. The first example is that of a one-dimensional $\sigma$-model with a symplectic target. The obstructions are then equivalent to those which must be overcome for a symplectic group action to admit an equivariant moment map. The other example is that of a two-dimensional WZW model with target space a Lie group admitting a bi-invariant metric. The obstruction in this case is simply an algebraic condition on the embedding of the symmetry algebra in the full isometry algebra.

In Section 7 we prove some vanishing theorems for $d$-dimensional $\sigma$-models for $d \leq 4$ under the assumption that the symmetry group is compact. These theorems guarantee the absence of obstructions for a variety of target manifolds. For example, it follows that any compact semisimple Lie group can be gauged in a three-dimensional $\sigma$-model with simply connected target space; or any compact semisimple Lie group having no cubic casimir can be gauged in a four-dimensional $\sigma$-model with target space another Lie group with at most one $U(1)$ factor. The compactness assumption on the symmetry group is necessary because of the topological methods we use, namely the Leray spectral sequence in singular cohomology. But since it is the algebraic de Rham model for equivariant cohomology which enters in the problem of gauging and not equivariant cohomology itself (although they agree for compact groups) the compactness assumption on the group should be inessential. Indeed we show that the vanishing persists *provided* that the target manifolds possess equivariant minimal models in the sense of rational homotopy theory. These are conjectural to the best of our knowledge, but we try to persuade the reader of their existence by constructing a few by hand. This may be of independent interest.

In Section 8 we recount of the main points and comment on possible extensions of the results in this paper. In particular we comment on the topological significance of the obstruction and on the entension to considering large gauge transformations.

We find it amusing to point out that we arrived at the conclusions in this paper almost by a slip of the tongue. During a seminar of Chris Hull on duality in string theory, he wrote down the conditions for gauging a two-dimensional $\sigma$-model. These conditions imply formally those for the existence of an equivariant moment mapping associated to the symplectic action of a Lie group; and in fact are equivalent to them in the case of gauging a one-dimensional $\sigma$-model. In Weinstein's lectures on symplectic geometry **[8]**, one can find an interpretation of these conditions in terms of "equivariant" cohomology; but upon closer examination it transpires that what Weinstein called "equivariant" in 1977 is however what today is termed invariant cohomology. Nevertheless, we remained with the attractive idea that (honest) equivariant cohomology would underpin these obstructions. The decisive hint was to be found in the paper of Atiyah and Bott **[9]**, where it is proven that the existence of a moment mapping is equivalent to the existence of a closed equivariant extension of the symplectic form.

Note added: During the process of typing the paper we became aware of Siye Wu's paper **[10]** where he treats the problem of finding a closed equivariant extension of a closed invariant form. In this paper he recovers the obstructions in the Lie algebraic formulation of Hull and Spence and he also comments on the topological interpretation in terms of the Leray spectral sequence. Moreover he made us aware of a paper of Edward Witten **[11]** containing an appendix which outlines already the relation between gauging a two-dimensional WZW model and equivariant cohomology. We are grateful to Siye Wu for his correspondence and for sending us a copy of his paper. The connection between equivariant cohomology and gauging (one-dimensional) $\sigma$-models is already implicit in work of George Papadopoulos **[12]**, predating **[11]**. We thank him for bringing this fact to our attention.

## §2 Gauging $\sigma$-Models

In this section we review briefly what we mean by gauging a $\sigma$-model and we set up the problem.

### Gauged $\sigma$-Models

The non-linear bosonic $\sigma$-model (without Wess–Zumino term) describes harmonic maps $\varphi : \Sigma \to M$ between two oriented (pseudo)riemannian manifolds $(\Sigma, h)$ and $(M, g)$. This latter manifold is referred to as the *target* manifold; whereas the $d$-dimensional manifold $\Sigma$ is called the *spacetime*; although we do not ascribe any definite signature to its metric. Let $\star$ denote the Hodge star operator on $\Sigma$ corresponding to the metric $h$ and the choice of orientation. The action describing the (principal) $\sigma$-model is

$$S_\sigma[\varphi] = \tfrac{1}{2}\int_\Sigma \|d\varphi\|^2 = \tfrac{1}{2}\int_\Sigma g_{ij}(\varphi) d\varphi^i \wedge \star d\varphi^j \ . \tag{2.1}$$

Its equations of motion are satisfied when $\varphi : \Sigma \to M$ is a harmonic map.

Now let $G$ be a connected Lie group acting on $M$ via isometries. Since it acts on $M$, it acts on the space $\mathrm{Map}(\Sigma, M)$ of maps from $\Sigma$ to $M$ in the obvious way and the $\sigma$-model action $S_\sigma$ is invariant under such transformations. Since $G$ is assumed connected, it is enough to check that the action is invariant under infinitesimal transformations. Let $\mathfrak{g}$ denote the Lie algebra of $G$ and for each $X \in \mathfrak{g}$ let $\xi_X$ denote the corresponding Killing vector field on $M$. Then one can check that under

$$\delta_X \varphi^i = \xi_X^i(\varphi) \tag{2.2}$$

the action transforms as

$$\delta_X S_\sigma[\varphi] = \tfrac{1}{2}\int_\Sigma \mathcal{L}_{\xi_X}\, g_{ij}(\varphi) d\varphi^i \wedge \star d\varphi^j \tag{2.3}$$

which vanishes by Killing's equation.

Gauging the $\sigma$-model consists of coupling it to gauge fields on $\Sigma$ in such a way that the resulting action is invariant under $\mathrm{Map}(\Sigma, G)$. We will only look at infinitesimal gauge transformations.

Infinitesimal gauge transformations are simply $\mathfrak{g}$-transformations with parameters which are functions on $\Sigma$. Explicitly, let us choose a basis $\langle X_a \rangle$ for $\mathfrak{g}$. Then an element $\lambda$ of $\mathrm{Map}(\Sigma, \mathfrak{g})$ can be written as $\lambda = \lambda^a X_a$ with $\lambda^a \in C^\infty(\Sigma)$. The infinitesimal gauge transformation with parameter $\lambda$ is given by

$$\delta_\lambda \varphi^i = \lambda^a \xi_a^i(\varphi) \ , \tag{2.4}$$

where $\xi_a$ is the Killing vector field associated to $X_a \in \mathfrak{g}$. It is easy to check that the $\sigma$-model action is no longer invariant under (2.4). To remedy this situation we introduce $\mathfrak{g}$-valued gauge fields $A = A^a X_a$, with $A^a$ locally defined one-forms on $\Sigma$ and transforming under infinitesimal gauge transformations as

$$\delta_\lambda A = d\lambda + [A, \lambda] \ . \tag{2.5}$$

This allows us to define the exterior covariant derivative

$$\nabla \varphi^i \equiv d\varphi^i - A^a \xi_a^i(\varphi) \ . \tag{2.6}$$

It is then easy to check that the new action

$$S_G[\varphi] = \tfrac{1}{2} \int_\Sigma \|\nabla\varphi\|^2 = \tfrac{1}{2} \int_\Sigma g_{ij}(\varphi) \nabla\varphi^i \wedge \star\nabla\varphi^j \tag{2.7}$$

is gauge-invariant. We will refer to the substitution of the exterior derivative by the exterior covariant derivative as *minimal coupling*.

<u>The Wess–Zumino Term</u>

Now consider a closed form $\omega \in \Omega^{d+1}(M)$. If $B$ is any $(d{+}1)$-manifold with boundary $\partial B = \Sigma$ and $\overline{\varphi} : B \to M$ extends $\varphi$ we can add to the $\sigma$-model action (2.1) the Wess–Zumino (WZ) term:

$$S_{\mathrm{WZ}}[\overline{\varphi}] = \int_B \overline{\varphi}^* \omega \ . \tag{2.8}$$

Because $d\omega = 0$, the variation of $S_{\mathrm{WZ}}$ is a boundary term and the resulting equations of motion depend only on the restriction of $\overline{\varphi}$ to $\Sigma$, namely on $\varphi$. In other words, the combined action $S = S_\sigma + S_{\mathrm{WZ}}$ defines a variational problem for maps $\varphi : \Sigma \to M$ which extend to a map $\overline{\varphi} : B \to M$ for some $(B, \overline{\varphi})$. The existence of an extension $\overline{\varphi} : B \to M$ given a $\varphi : \Sigma \to M$ can be analyzed via topological obstruction theory. Demanding that this be true for all $\varphi$ but for fixed $\Sigma$ one finds that certain topological cohomology spaces have to vanish. We will not consider these obstructions in this paper. Furthermore, since our considerations are purely classical, neither will we bother with the integrality conditions placed on $\omega$ to guarantee that the quantum physics defined by the WZ term is independent of the extension. We will henceforth assume that $S_{\mathrm{WZ}}$ exists and in well-defined and we will only address the obstructions to gauging it.

Assume now that $\omega$ is $G$-invariant. Then so is the WZ term and we can address the question of whether $S_{\mathrm{WZ}}$ can be promoted to a gauge-invariant action. Unlike the case with the action (2.1), minimal coupling does not suffice, for whereas it does produce a gauge-invariant action, the equations of motion now do depend on the extension. Indeed, if we minimally couple $\overline{\varphi}^*\omega$ it is easy to show explicitly that its contribution to the equations of motion are however dependent on the extension $\overline{\varphi} : B \to M$. This is due to the fact that the minimally coupled $\overline{\varphi}^*\omega$ is not closed.

The question we address in this paper is the following. ***Under what conditions can $\overline{\varphi}^*\omega$ be modified in such a way that the resulting action is gauge-invariant, defines "local" (that is, independent of the extension) equations of motion, and reduces to $S_{\mathrm{WZ}}$ when the gauge fields are put to zero?***

Notice that if $\omega = d\theta$ and $\theta$ is $G$-invariant, then

$$S_{\mathrm{WZ}}[\overline{\varphi}] = \int_B \overline{\varphi}^* d\theta = \int_\Sigma \varphi^*\theta \ ; \tag{2.9}$$

and this can be gauged by minimal coupling just like $S_\sigma$. This condition, while certainly sufficient, is clearly not necessary and we will devote the rest of the paper to studying the precise conditions under which a given $S_{\mathrm{WZ}}$ can be gauged. We will see that this problem can be phrased very naturally in terms of the equivariant cohomology of the target manifold.

Topological Heuristics

We can understand a priori why this would have to be the case by looking at the problem from a slightly different perspective. Rather than thinking about gauging a $\sigma$-model, we can take the point of view that the gauge fields are the fundamental objects of the theory and we can ask ourselves the question of when can we couple them to matter fields in the form of a nonlinear $\sigma$-model.

The gauge fields are described geometrically by a connection in a principal $G$-bundle $P \to \Sigma$, which in preparation to our discussion of the Wess–Zumino term, can be thought of as the restriction to $\Sigma = \partial B$ of a principal $G$-bundle $\overline{P} \to B$. Matter fields are then sections through some associated bundle. The usual matter fields in renormalizable quantum field theory are in a linear representation of $G$; that is, the associated bundle is a vector bundle. In the case of the $\sigma$-model, however, the group is nonlinearly realized as diffeomorphisms of a manifold $M$. In other words, the matter fields in a gauged $\sigma$-model will be sections of as associated bundle $P \times_G M \to \Sigma$, which will be the restriction to $\Sigma$ of the bundle $\overline{P} \times_G M \to B$. Contrary to vector bundles—which always have sections—generic fiber bundles may have none. We will assume for simplicity that our bundle is such that sections do exist; for example, it could be the trivial bundle, where a section is simply a map $\Sigma \to M$.

By universality, the bundle $\overline{P} \to B$ is the pull-back bundle of the universal bundle $EG \to BG$ via some map $\overline{\gamma} : B \to BG$ which is well-defined up to homotopy. Similarly, the associated bundle $\overline{P} \times_G M \to B$ is the pullback under the same map of the bundle $EG \times_G M \to BG$. In other words, we have the commutative diagram

$$\begin{array}{ccccc} P \times_G M & \longrightarrow & \overline{P} \times_G M & \xrightarrow{\widetilde{\gamma}} & EG \times_G M \\ \downarrow & & \downarrow & & \downarrow \\ \Sigma & \longrightarrow & B & \xrightarrow{\overline{\gamma}} & BG \end{array} \tag{2.10}$$

We shall use the common abbreviation $M_G$ for $EG \times_G M$. A section $\overline{\varphi} : B \to \overline{P} \times_G M$ defines a map $\widetilde{\gamma} \circ \overline{\varphi} : B \to M_G$ which allows us to pull forms back from $M_G$ to $B$. In particular, the integral over $B$ of the pull-back of a closed $(d+1)$-form on $M_G$ gives rise to a topological term for the gauged $\sigma$-model; that is, a term which although defines a variational problem on $\Sigma$ cannot be defined locally on $\Sigma$ but rather on $B$. Of course, if the form is exact, then the term is an integral over $\Sigma$ and hence does not correspond to a topological term. Hence we see that topological terms in a gauged $\sigma$-model are in one-to-one correspondence with $H^{d+1}(M_G)$. Now, by definition the cohomology of $M_G$ is the $G$-equivariant cohomology of $M$. Hence topological terms for a gauged $\sigma$-model correspond precisely with $G$-equivariant cohomology classes for $M$. It is interesting to compare this with Witten's observation **[13]** that the physical spectrum of a gauged topological $\sigma$-model coincides with the $G$-equivariant cohomology of $M$. Hence there is an intriguing one-to-one correspondence between the spectrum of a topological $\sigma$-model coupled to Yang-Mills and the topological terms of a bosonic $\sigma$-model coupled to Yang-Mills.

## §3 Gauging and Equivariant Cohomology

Having seen that equivariant cohomology plays a role in gauging Wess–Zumino terms, we exhibit the precise relation in this section. In fact, what we will see is that there is an intimate relation between gauged Wess–Zumino terms and cohomology classes in the (algebraic) de Rham model for $G$-equivariant cohomology. This model agrees with the topological definition of equivariant cohomology for $G$ compact, but computes something else altogether for noncompact $G$. In fact as will become clear below, it is the Lie algebra $\mathfrak{g}$ of $G$ that plays the important role, and as usual one does not expect that for noncompact Lie groups one can get any kind of topological information by localization. We start by briefly reviewing the differential graded algebra associated to a connection in a principal fiber bundle and then introduce the Weil algebra as a universal model for such situations. This allows us to use the Weil algebra in a de Rham model for equivariant cohomology. We will see that the gauge-invariant forms correspond precisely to the basic (or equivariant) forms. This immediately allows us to interpret the obstruction to gauging a given Wess–Zumino term as the obstruction to finding a closed equivariant extension of an invariant form on $M$. *Throughout this paper we shall take $G$ to be a connected Lie group with Lie algebra $\mathfrak{g}$.*

### The Differential Graded Algebra of a Principal Bundle with Connection

Let $\pi : E \to X$ be any principal $G$-bundle. In particular, $G$ acts freely on $E$ on the right, and $X \cong E/G$. The action of $G$ induces an infinitesimal action of the Lie algebra $\mathfrak{g}$. If $X \in \mathfrak{g}$, we let $\xi_X$ denote the corresponding vector field on $E$. This maps sets up an isomorphism $\mathfrak{g} \to V_e$ for all $e \in E$, where $V \subset TE$ is the vertical subbundle defined by $V = \ker \pi_*$. Let $H \subset TE$ be a connection and let $\omega \in \Omega^1(E) \otimes \mathfrak{g}$ be the $\mathfrak{g}$-valued connection one-form. Recall that $\omega$ annihilates $H$ and $\omega(\xi_X) = X$. Let $\phi \in \Omega^2(E) \otimes \mathfrak{g}$ be its curvature two-form. It is defined by the structure equation

$$\phi = d\omega + \tfrac{1}{2}[\omega, \omega] \ . \tag{3.1}$$

Applying $d$ to both sides one gets the Bianchi identity

$$d\phi = [\phi, \omega] \ . \tag{3.2}$$

Relative to a fixed basis $\langle X_a \rangle$ for $\mathfrak{g}$, we expand $\omega = \omega^a X_a$ and $\phi = \phi^a X_a$. Equations (3.1) and (3.2) become

$$\phi^a = d\omega^a + \tfrac{1}{2} f_{bc}{}^a \omega^b \omega^c \qquad \text{and} \qquad d\phi^a = f_{bc}{}^a \phi^b \omega^c \ , \tag{3.3}$$

which show that the subalgebra of $\Omega(E)$ generated by $\omega^a$ and $\phi^a$ is closed under $d$. In other words, this makes it into a differential graded subalgebra of $\Omega(E)$. On $\Omega(E)$ we have associated to each $X \in \mathfrak{g}$ an antiderivation $\imath(\xi_X)$. By definition $\imath(\xi_X)\omega = X$, and it follows from (3.1) that $\imath(\xi_X)\phi = 0$. In general, any form in $\Omega(E)$ which is annihilated by $\imath(\xi_X)$ for all $X$ is called ***horizontal***. We now define an action of $\mathfrak{g}$ via the Lie derivative $\mathcal{L}_X = [d, \imath(\xi_X)]$. One checks that $[\mathcal{L}_X, \imath(\xi_Y)] = \imath(\xi_{[X,Y]})$ and that $[\mathcal{L}_X, \mathcal{L}_Y] = \mathcal{L}_{[X,Y]}$. A form on $\Omega(E)$ is called ***invariant*** if it is annihilated by $\mathcal{L}_X$ for all $X \in \mathfrak{g}$. It is easy to see that

$$\mathcal{L}_X\omega = -[X, \omega] \qquad \text{and} \qquad \mathcal{L}_X\phi = -[X, \phi] \; ; \tag{3.4}$$

or relative to the basis $\langle X_a \rangle$ for $\mathfrak{g}$, and letting $\imath_a$ and $\mathcal{L}_a$ stand for $\imath(\xi_{X_a})$ and $\mathcal{L}_{X_a}$ respectively,

$$\mathcal{L}_a\omega^c = -f_{ab}{}^c\omega^b \qquad \text{and} \qquad \mathcal{L}_a\phi^c = -f_{ab}{}^c\phi^b \; . \tag{3.5}$$

Clearly $\mathcal{L}_X$ commutes with $d$ hence it induces an action in cohomology. However the action is trivial since $\mathcal{L}_X$ is (manifestly) null homotopic. Following Cartan **[14]**, we find it convenient to give this structure a name.

DEFINITION 3.6. By a $\mathfrak{g}$-*DGA* we mean a differential graded (commutative) algebra (DGA) $(A, d)$ with an null-homotopic action of $\mathfrak{g}$ given by $X \mapsto \mathcal{L}_X \equiv [d, \imath_X]$, where $\imath_X$ are degree $-1$ antiderivations satisfying $[\mathcal{L}_X, \imath_Y] = \imath_{[X,Y]}$.

In other words, for any principal $G$-bundle $E \to X$, $\Omega(E)$ is a $\mathfrak{g}$-DGA and, having specified a connection, the subalgebra generated by $\omega^a$ and $\phi^a$ is a sub-$\mathfrak{g}$-DGA.

The projection $\pi : E \to X$ defines an embedding of differential graded algebras $\pi^* : \Omega(X) \hookrightarrow \Omega(E)$. It is easy to check that the image of $\pi^*$ contains those forms on $E$ which are horizontal and invariant; and it is not hard to prove that it does not contain anything else. Consequently those forms in $\Omega(E)$ which are both horizontal and invariant are called ***basic*** forms. The basic forms form a subcomplex of $\Omega(E)$ isomorphic to $\Omega(X)$, and hence its cohomology coincides with the de Rham cohomology of the base $X$. Let us remark that the notions of horizontal, invariant, and basic forms make sense for any $\mathfrak{g}$-DGA even if it is not the $\mathfrak{g}$-DGA of a principal $G$-bundle.

The Weil Algebra

In the late 1940's André Weil (see **[14]**) introduced a universal model for the $\mathfrak{g}$-DGA of a principal $G$-bundle. The Weil algebra $W(\mathfrak{g})$ is defined by

$$W(\mathfrak{g}) \equiv \bigwedge\mathfrak{g}^* \otimes \mathfrak{S}\mathfrak{g}^* \; , \tag{3.7}$$

where $\bigwedge\mathfrak{g}^*$ and $\mathfrak{S}\mathfrak{g}^*$ are the exterior and symmetric algebras of $\mathfrak{g}^*$. Choose a basis $\langle \theta^a \rangle$ for $\mathfrak{g}^*$ canonically dual to the basis $\langle X_a \rangle$ for $\mathfrak{g}$. The element corresponding to $\theta^a$ in $\bigwedge^1\mathfrak{g}^*$ will be denoted $A^a$, whereas the one in $\mathfrak{S}^1\mathfrak{g}^*$ will be denoted $F^a$. The Weil algebra is graded by demanding that $A^a$ have degree 1 and $F^a$ have degree 2. We define a differential $d$ on $W(\mathfrak{g})$ by

$$\begin{aligned} dA^a &= -\tfrac{1}{2} f_{bc}{}^a A^b A^c + F^a \\ dF^a &= f_{bc}{}^a F^b A^c \end{aligned} \tag{3.8}$$

and extending it to all of $W(\mathfrak{g})$ as an antiderivation. It is easy to verify that $d^2 = 0$ so that it does defines a differential graded algebra. We can think of $A^a$ and $F^a$ as the components relative to the basis $\langle X_a\rangle$ for $\mathfrak{g}$ of $\mathfrak{g}$-valued forms $A = A^a X_a \in \bigwedge^1\mathfrak{g}^* \otimes \mathfrak{g}$ and $F = F^a X_a \in \mathfrak{S}^1\mathfrak{g}^* \otimes \mathfrak{g}$. We now define antiderivations $\imath_X$ of degree $-1$ for each $X \in \mathfrak{g}$ by $\imath_X A = X$, and $\imath_X F = 0$. We define an action of $\mathfrak{g}$ on $W(\mathfrak{g})$ by the degree 0 derivations $\mathfrak{L}_X = [d, \imath_X]$. One can prove from the definitions that $[\mathfrak{L}_X, \imath_Y] = \imath_{[X,Y]}$ and that $[\mathfrak{L}_X, \mathfrak{L}_Y] = \mathfrak{L}_{[X,Y]}$; whence the Weil algebra $W(\mathfrak{g})$ is a $\mathfrak{g}$-DGA.

Since the equations (3.8) are precisely (3.3), we have a map $w : W(\mathfrak{g}) \to \Omega(E)$, for any principal $G$-bundle $E \to X$. It is defined by $w(A) = \omega$ and $w(F) = \phi$, and $w \circ \imath_X = \imath(\xi_X) \circ w$. One can verify that $w$ defined this way is a morphism of $\mathfrak{g}$-DGAs. It is called the *Weil homomorphism*.

On the other hand, every principal $G$-bundle $E \to X$ is the pull-back of the universal bundle $EG \to BG$ via a map $\gamma : X \to BG$, which is well-defined up to homotopy. That is, we have a $G$-bundle map:

$$\begin{array}{ccc} E & \xrightarrow{\widetilde{\gamma}} & EG \\ \downarrow & & \downarrow \\ X & \xrightarrow{\gamma} & BG \end{array} \tag{3.9}$$

If $EG$ were a manifold, this would induce a map of $\mathfrak{g}$-DGAs $\widetilde{\gamma}^* : \Omega(EG) \to \Omega(E)$. In fact, $EG$ is not a manifold; although it is an inductive limit of manifolds and hence the above remarks allow us to understand the Weil algebra as a de Rham model for $EG$, at least for compact $G$. Indeed, the cohomology the Weil algebra is trivial, as befits a contractible space like $EG$; and also the cohomology of the basic subcomplex of $W(\mathfrak{g})$ agrees (for $G$ compact) with the cohomology of the classifying space $BG$.

### Equivariant Cohomology

Let $M$ be a manifold on which a connected compact Lie group $G$ acts. If the action of $G$ were free, the quotient $M/G$ would be a manifold. However for arbitrary group actions, $M/G$ is not even guaranteed to be Hausdorff. To remedy this fact, topologists have come up with a construction known as the ***homotopy quotient***. This construction is better behaved and agrees (up to homotopy) with the topological quotient $M/G$ when the action is free. The construction is the following. Consider $EG \times M$. Since $G$ acts freely on $EG$, it acts freely on $EG \times M$. Define $EG \times_G M \equiv (EG \times M)/G$. Because the action of $G$ on $EG$ is fiberwise, there is a well-defined map $\pi : EG \times_G M \to BG$ defined by sending the orbit through $(e, m)$ in $EG \times_G M$ to the orbit through $e$ in $BG$. This map is indeed a fibration and the typical fiber is easily seen to be $M$. Similarly, if the action of $G$ on $M$ is free, there is a well-defined map $EG \times_G M \to M/G$ which is also a fibration with typical fiber $EG$. Since $EG$ is contractible, $EG \times_G M$ has the homotopy type of $M/G$. In general, we have the diagram

$$\begin{array}{ccccc} EG & \longleftarrow & EG \times M & \longrightarrow & M \\ \downarrow & & \downarrow & & \downarrow \\ BG & \xleftarrow{\pi} & EG \times_G M & \xrightarrow{\sigma} & M/G \end{array} \tag{3.10}$$

where $\pi$ is a fibration but where $\sigma$ is not unless the action is free.

The homotopy quotient $EG \times_G M$ is often abbreviated $M_G$. The equivariant cohomology of $M$ is then defined as the (singular) cohomology of $M_G$ and it is denoted $H_G(M)$. It is naturally a module over $H(BG)$, the module structure being induced by the map $\pi^* : H(BG) \to H(M_G)$. By the homotopy invariance of cohomology, in the case of a free action, $H_G(M) \cong H(M/G)$. At the other extreme, if $M$ is a point, then $H_G(\mathrm{pt}) \cong H(BG)$ which is huge.

Using the Weil algebra we can give a de Rham model for equivariant cohomology. Under the identification of $W(\mathfrak{g})$ and $\Omega(EG)$, we can think of $\Omega(EG \times M)$ as modeled by $W(\mathfrak{g}) \otimes \Omega(M)$. Notice that since $G$ acts on $M$, $\Omega(M)$ becomes a $\mathfrak{g}$-DGA and so does the tensor product $W(\mathfrak{g}) \otimes \Omega(M)$. The basic subalgebra $\Omega_{\mathfrak{g}}(M)$ of $W(\mathfrak{g}) \otimes \Omega(M)$ constitutes a model for the basic forms on $EG \times M$; that is, for the forms on $M_G$. Analogously to the topological situation described above, one defines the equivariant cohomology of $\Omega(M)$ as the cohomology of the basic subcomplex of $W(\mathfrak{g}) \otimes \Omega(M)$. This definition makes sense for all Lie algebras $\mathfrak{g}$ (as long as $\Omega(M)$ is a $\mathfrak{g}$-DGA); but only for compact $G$ will the equivariant cohomology of $\Omega(M)$ agree with the equivariant cohomology of $M$ defined in the previous paragraph.

For convenience we record the following algebraic definition of equivariant cohomology:

Definition 3.11. Let $(A, d)$ be any $\mathfrak{g}$-DGA and let $A_{\mathfrak{g}}$ denote the subalgebra of basic forms in $W(\mathfrak{g}) \otimes A$. The *equivariant cohomology* of $A$, denoted $H_{\mathfrak{g}}(A)$ is defined as the cohomology $H(A_{\mathfrak{g}})$ of the basic subalgebra. If $M$ is a manifold, then $H_{\mathfrak{g}}(M) \equiv H_{\mathfrak{g}}(\Omega(M))$.

Gauge Transformations in a $\sigma$-Model and Basic Forms

We consider the $\mathfrak{g}$-DGA $W(\mathfrak{g}) \otimes \Omega(M)$. Its elements, which we shall call "forms" from now on, are finite linear combinations of monomials of the form

$$\omega_{\underbrace{a\cdots b}_{p}\underbrace{c\cdots d}_{q}} A^a \cdots A^b F^c \cdots F^d \ , \tag{3.12}$$

where $\omega_{a\cdots bc\cdots d} \in \Omega^r(M)$. We grade $W(\mathfrak{g}) \otimes \Omega(M)$ with the total degree coming from its factors so that, for example, the monomial in (3.12) has degree $p + 2q + r$.

We will now show that the gauge-invariant forms of the $\sigma$-model with target space $M$ are precisely the basic forms on $W(\mathfrak{g}) \otimes \Omega(M)$; thus substantiating the topological heuristics of the last section.

In order to discuss gauge transformations in this abstract setting, we find it necessary to introduce an auxiliary differential graded algebra $\mathcal{E}$ with generators $\lambda^a$ of degree 0 and $d\lambda^a$ of degree 1. Gauge transformations can be understood as a morphism of DGAs

$$\delta_\lambda : W(\mathfrak{g}) \otimes \Omega(M) \longrightarrow \mathcal{E} \otimes W(\mathfrak{g}) \otimes \Omega(M) \tag{3.13}$$

defined by

$$\delta_\lambda \phi = d\lambda^a \imath_a \phi + \lambda^a \mathcal{L}_a \phi \ . \tag{3.14}$$

On the generators of the Weil algebra, we recover the familiar formulas

$$\begin{aligned} \delta_\lambda A^a &= d\lambda^a - f_{bc}{}^a \lambda^b A^c \\ \delta_\lambda F^a &= -f_{bc}{}^a \lambda^b F^c \end{aligned} \tag{3.15}$$

whereas on $\Omega(M)$, (3.14) mimics the transformation law of the pullback to $B$ of a form on $M$ under (2.4), namely

$$\delta_\lambda \overline{\varphi}^* \omega = d\lambda^a \wedge \overline{\varphi}^* \imath_a \omega + \lambda^a \overline{\varphi}^* \mathcal{L}_a \omega \ . \tag{3.16}$$

It follows at once from (3.14) that the gauge invariant forms in $W(\mathfrak{g}) \otimes \Omega(M)$ are precisely the basic forms $\Omega_{\mathfrak{g}}(M)$! In view of this fact, we can reinterpret the problem of gauging the WZ term as follows. Given a closed $G$-invariant form $\omega \in \Omega(M)$, we want to extend it, by adding pieces depending on $A^a$ and $F^a$, to a gauge-invariant form on $W(\mathfrak{g}) \otimes \Omega(M)$ which is still closed, so that its contribution to the equations of motion are local. A closed equivariant form which reduces to $\omega$ when $A^a$ and $F^a$ are put to zero will be called a closed equivariant extension of $\omega$. Therefore we have proven the following result, which forms the basis of the rest of this paper.

PROPOSITION 3.17. *The Wess–Zumino term* (2.8) *can be gauged if and only if $\omega$ has an equivariant closed extension.*

We should emphasize that despite the topological coherence of the construction, the problem at hand is algebraic. It is the equivariant extension in the sense of the algebraic de Rham model for equivariant cohomology, and not its topological counterpart, that is equivalent to gauging. Therefore when $G$ is not compact, the obstruction will not really be directly related with the topology of the homotopy quotient $M_G$.

## §4 Universal Obstructions

In this section we analyze the obstructions to gauging the WZ term in the new light shed by Proposition 3.17. Namely, we will analyze the obstructions to finding a closed equivariant extension of a given invariant closed form $\omega$ on $M$. We will see that the obstruction is cohomological and that it defines a class in a subcomplex of the equivariant forms on $M$. This subcomplex fits in a short exact sequence of complexes and the obstruction is nothing but the image under the connecting homomorphism of the class of $\omega$ in the $\mathfrak{g}$-invariant cohomology of $M$.

This way of describing the obstruction suggests a method of gauging which differs from the more traditional Noether procedure. In the Noether procedure one starts with $\omega$ which is closed but not equivariant and tries to extend it to an equivariant form by adding closed (indeed, exact) pieces depending on $A$ and $F$. The alternate method suggested by equivariant cohomology is different. There we first minimally couple $\omega$ to make it equivariant and then we subtract equivariant pieces to try to close the resulting form. A closer look at the Noether procedure reveals that the obstruction to gauging can be understood as a class living in a different cohomology theory, namely BRST cohomology. Indeed the obstruction is a class in BRST cohomology at ghost number 1. This is reminiscent of the homological interpretation of anomalies and we comment briefly on the connection.

We should remark that since we are working with universal objects instead of with the original bundle $\overline{P} \to B$, the obstructions we will discover are universal. That is, their vanishing will be a sufficient condition for the gauging to be possible, but it may be that for particular choices of $\overline{P} \to B$, some additional structure may allow the cancellation of the obstruction after pulling the form back.

Let us elaborate on this point. If $\overline{\varphi} : B \to M$ is any map, we have a map $\overline{\varphi}^* : \Omega(M) \to \Omega(B)$. Furthermore we can compose this map with the pullback of the projection $\overline{\pi} : \overline{P} \to B$ to obtain a map $\overline{\mu}^* \equiv (\overline{\varphi} \circ \overline{\pi})^* : \Omega(M) \to \Omega(\overline{P})$. Combining $\overline{\mu}^*$ with the Weil homomorphism $w : W(\mathfrak{g}) \to \Omega(\overline{P})$, we have a map $w \otimes \overline{\mu}^* : W(\mathfrak{g}) \otimes \Omega(M) \to \Omega(\overline{P})$. This map clearly maps basic forms to basic forms. Hence if $\omega^\#$ is a closed basic form in $W(\mathfrak{g}) \otimes \Omega(M)$, so will $(w \otimes \overline{\mu}^*)(\omega^\#)$ be. It is in fact this latter form that we will use to define the gauged WZ action, but we will be analyzing the obstructions in $W(\mathfrak{g}) \otimes \Omega(M)$ before mapping back to $\Omega(\overline{P})$, hence their universality.

## An Exact Sequence and the Obstruction

Let us denote by $\Omega(M)^{\mathfrak{g}}$ the subalgebra of $\Omega(M)$ consisting of the invariant forms:

$$\Omega(M)^{\mathfrak{g}} \equiv \{\omega \in \Omega(M) \mid \mathcal{L}_X\, \omega = 0 \;\forall\; X \in \mathfrak{g}\} \;. \tag{4.1}$$

Because $\mathcal{L}_X$ commutes with $d$, $\Omega(M)^{\mathfrak{g}}$ is a subcomplex of the de Rham complex on $M$ and we will let $H(M)^{\mathfrak{g}}$ denote its cohomology. This is the $\mathfrak{g}$-invariant cohomology of $M$.

Given any equivariant form $\omega^\# \in \Omega_{\mathfrak{g}}(M)$, the form obtained putting $A^a{=}F^a{=}0$ is a $\mathfrak{g}$-invariant form $\omega \in \Omega(M)^{\mathfrak{g}}$. Furthermore, this map commutes with $d$ so that we have a natural morphism of complexes

$$\Phi : \Omega_{\mathfrak{g}}(M) \longrightarrow \Omega(M)^{\mathfrak{g}} \;. \tag{4.2}$$

This map is actually surjective. Indeed, given a $\mathfrak{g}$-invariant form $\omega \in \Omega(M)^{\mathfrak{g}}$ we can always extend it to an equivariant form by minimal coupling. (See later for a more explicit description of minimal coupling.) Now let $\Omega_\Phi(M) \subset \Omega_{\mathfrak{g}}(M)$ denote the kernel of $\Phi$. It is of course a subcomplex since it is the kernel of morphism, and it is explicitly given by those equivariant forms which vanish when $A^a{=}F^a{=}0$. We can rephrase these two observations by saying that we have a short exact sequence of graded complexes:

$$0 \longrightarrow \Omega_\Phi(M) \longrightarrow \Omega_{\mathfrak{g}}(M) \xrightarrow{\;\Phi\;} \Omega(M)^{\mathfrak{g}} \longrightarrow 0 \;. \tag{4.3}$$

Naturally, this induces a long exact sequence in cohomology:

$$\begin{array}{l} \cdots \\ \longrightarrow H_\Phi^{d+1}(M) \longrightarrow H_{\mathfrak{g}}^{d+1}(M) \xrightarrow{\;\Phi_*\;} H^{d+1}(M)^{\mathfrak{g}} \xrightarrow{\;d_*\;} \\ \longrightarrow H_\Phi^{d+2}(M) \longrightarrow \cdots \end{array} \tag{4.4}$$

where $H_\Phi(M)$ denotes the cohomology of $\Omega_\Phi(M)$, and $d_*$ denotes the connecting homomorphism.

Now let $\omega \in \Omega^{d+1}(M)$ be the closed $\mathfrak{g}$-invariant form defining the Wess–Zumino term. It defines a class in $H^{d+1}(M)^{\mathfrak{g}}$. By Proposition 3.17, the Wess–Zumino term can be gauged if $\omega = \Phi(\omega^\#)$ for some equivariant closed form $\omega^\# \in \Omega_{\mathfrak{g}}^{d+1}(M)$. In other words, the Wess–Zumino term can be gauged if and only if the cohomology class defined by $\omega$ lies in the image of $\Phi_*$. By the exactness of (4.4), this coincides with the kernel of the connecting homomorphism $d_*$; hence we have the following result:

THEOREM 4.5. *The (universal) obstruction to gauging the Wess–Zumino term defined by* $\omega \in \Omega^{d+1}(M)^{\mathfrak{g}}$ *is the class* $d_*[\omega] \in H_\Phi^{d+2}(M)$.

Notice that the obstruction trivially vanishes when the class of $\omega$ in $H(M)^{\mathfrak{g}}$ is trivial. In this case $\omega = d\theta$ for $\theta$ a $\mathfrak{g}$-invariant form on $M$. But as mentioned earlier, the WZ term then becomes local and can be gauged trivially via minimal coupling. To arrive at some nontrivial corollaries of this theorem, it will be necessary to take a closer look at $H_\Phi(M)$ and in particular at the class $d_*[\omega]$. We postpone the former problem until a later section and we concentrate on the latter. To elucidate the nature of the class $d_*[\omega]$, it will be necessary to understand the definition of the connecting homomorphism $d_*$ in (4.4).

Take $\omega \in \Omega^{d+1}(M)^{\mathfrak{g}}$ a closed invariant form and let $\widetilde{\omega}$ denote the equivariant form obtained from $\omega$ by minimal coupling. $\widetilde{\omega}$ will fail in general to be closed—this is precisely the nature of the obstruction. But since $\omega$ is closed, $\Phi(d\widetilde{\omega}) = d\Phi(\widetilde{\omega}) = d\omega = 0$. In other words, by the exactness of (4.3), $d\widetilde{\omega} \in \Omega_\Phi^{d+2}(M)$. It is obviously closed, but not necessarily exact since $\widetilde{\omega} \notin \Omega_\Phi(M)$. The class $[d\widetilde{\omega}] \in H_\Phi^{d+2}(M)$ is precisely $d_*[\omega]$. This class will vanish if there exists a $\widetilde{\Delta} \in \Omega_\Phi^{d+1}(M)$ such that $d\widetilde{\omega} = d\widetilde{\Delta}$. If this is so, $\omega^\# \equiv \widetilde{\omega} - \widetilde{\Delta}$ is closed, equivariant and, since $\Phi(\omega^\#) = \Phi(\widetilde{\omega}) = \omega$, it extends $\omega$.

## A Cohomological Look at the Noether Method

Let us briefly contrast the Noether method to gauge the WZ term $\omega$ from that suggested by Theorem 4.5 and discussed in the previous paragraph. The Noether method starts with $\omega$, which is closed but not gauge-invariant. Indeed, its gauge variation $\delta_\lambda \omega$ will not be zero in general; but it will be independent of the gauge field $A$. The first step is to add a *closed* term $\omega_{(1)}$ linear in $A$ and $dA$ (which, instead of introducing $F$, we leave as $dA$ for the present purposes) to cancel the gauge variation of $\omega$ up to terms linear in $A$, where we follow the convention that $dA$ is linear in $A$. Cancelling the variation is easy, but doing it via a closed form is not. This yields an obstruction. If this obstruction is overcome, we go on to the next step: we add another closed form $\omega_{(2)}$ quadratic in $A$ in such a way that the gauge variation of $\omega + \omega_{(1)} + \omega_{(2)}$ is quadratic in $A$. Again there may be an obstruction to doing this. Continuing in this fashion we arrive at a form $\omega + \omega_{(1)} + \omega_{(2)} + \cdots = \omega - \Theta$ which is gauge-invariant and closed. In other words, the Noether procedure consists in finding a closed form $\Theta$ which vanishes when $A$=$F$=0 and such that $\omega - \Theta$ is gauge-invariant. Notice that since $\omega - \Theta$ is in particular invariant under constant gauge transformations and since $\omega$ is separately invariant , then so is $\Theta$. In summary, the Noether procedure consists in finding a $\mathfrak{g}$-invariant, closed form $\Theta$ which vanishes when $A$=$F$=0 and such that $\delta_\lambda \Theta = \delta_\lambda \omega$.

We should contrast this with the treatment of the obstructions found in the previous subsection. There we first minimally couple $\omega$ to a gauge-invariant form $\widetilde{\omega}$ and then asked for the existence of a gauge invariant form $\widetilde{\Delta}$ which vanishes when $A = F = 0$ and such that $d\widetilde{\Delta} = d\widetilde{\omega}$. This description suggested a homological interpretation culminating in Theorem 4.5. We would now like to re-interpret the obstructions homologically in the manner suggested by the Noether procedure.

The formal similarity between the two approaches suggest that we exchange $d$ and $\delta_\lambda$. Since $\delta_\lambda$ is not a differential, we will need to replace it by the BRST operator. The cohomology theory in which the obstruction to gauging will be seen to live turns out to be related to the one where consistent anomalies live. On the one hand, this should come as no surprise given the intimate relation between consistent anomalies and the WZW model; but on the other hand, it may be somewhat surprising that the relation persists even when the WZ term has no clear "anomalous" interpretation.

The first step is to introduce a differential graded algebra $\mathcal{G}$ generated by objects $c^a$ and $dc^a$. It is analogous to the auxiliary DGA $\mathcal{E}$ in (3.13), except for a regrading of the generators $c^a$. Indeed, we grade $\mathcal{G}$ by ***ghost number***, which is defined on generators by declaring that $c^a$ and $dc^a$ both have ghost number 1 and extending it additively to monomials. We extend the notion of ghost number to $\mathcal{C} \equiv W(\mathfrak{g}) \otimes \Omega(M) \otimes \mathcal{G}$ by declaring it to be zero on $W(\mathfrak{g}) \otimes \Omega(M)$. The notion of form-degree can similarly be extended to $\mathcal{C}$ by declaring that $c^a$ has form-degree 0 and $dc^a$ has form-degree 1. The total degree (which determines the odd-even parity of the objects) is given by the sum of the form-degree and the ghost number. Hence, $c^a$ has total degree 1, whereas $dc^a$ has total degree 2. On $W(\mathfrak{g}) \otimes \Omega(M)$ the total degree agrees with the form-degree. In summary, $\mathcal{C}$ is bigraded by form-degree and ghost number $\mathcal{C} = \bigoplus_{p,q\geq 0} \mathcal{C}^{p,q}$, where $\mathcal{C}^{p,q}$ is the subspace of $\mathcal{C}$ consisting of forms of form-degree $p$ and ghost number $q$. If need be, we will refer to the form-degree simply as degree and to the total degree as the dimension.

Relative to the differential $d : \mathcal{C}^{p,q} \to \mathcal{C}^{p+1,q}$, $\mathcal{C}$ becomes a DGA whose cohomology clearly agrees with the de Rham cohomology of $M$. We can extend the action of $G$ from $W(\mathfrak{g}) \otimes \Omega(M)$ to $\mathcal{C}$ by defining the action of $G$ on $\mathcal{G}$ as follows. We demand that both $c = c^a X_a \in \mathcal{G} \otimes \mathfrak{g}$ and $dc = dc^a X_a \in \mathcal{G} \otimes \mathfrak{g}$ be $\mathfrak{g}$-invariant. In other words, the action of $\mathfrak{g}$ on $\mathcal{G}$ is given by

$$\mathcal{L}_a\, c^b = -f_{ac}{}^b c^c \qquad \text{and} \qquad \mathcal{L}_a\, dc^b = -f_{ac}{}^b dc^c \ , \tag{4.6}$$

which can be written in the $\mathfrak{g}$-DGA form $\mathcal{L}_a = [d, \imath_a]$, provided we define the antiderivations $\imath_a$ by

$$\imath_a c^b = 0 \qquad \text{and} \qquad \imath_a dc^b = -f_{ac}{}^b c^c \ . \tag{4.7}$$

It is also easy to show that the $\mathfrak{g}$-invariant cohomology of $\mathcal{C}$ coincides with the $\mathfrak{g}$-invariant cohomology of $M$.

The BRST operator $\delta : \mathcal{C}^{p,q} \to \mathcal{C}^{p,q+1}$ is defined as follows. On $\phi \in W(\mathfrak{g}) \otimes \Omega(M)$ it mimics the gauge transformation (3.14) (but notice the relative sign!)

$$\delta\phi = c^a\, \mathcal{L}_a\, \phi - dc^a \imath_a \phi \ ; \tag{4.8}$$

whereas on $c^a$ it is defined in such a way that $\delta^2 = 0$; namely,

$$\delta c^a = -\tfrac{1}{2} f_{bc}{}^a c^b c^c \ . \tag{4.9}$$

We extend it to all of $\mathcal{C}$ as an antiderivation anticommuting with $d$. In summary $(\mathcal{C}^{\bullet,\bullet}, d, \delta)$ becomes a double complex. Its associated total complex is defined as usual by $\mathcal{C}^n = \bigoplus_{p+q=n} \mathcal{C}^{p,q}$ and $D = d + \delta$; whence $D : \mathcal{C}^n \to \mathcal{C}^{n+1}$ and $(\mathcal{C}^\bullet, D)$ becomes a graded complex.

It is easy to prove that the cohomology of the total complex is isomorphic to the de Rham cohomology of $M$, whereas the cohomology of the $\mathfrak{g}$-invariant subcomplex, is isomorphic to the $\mathfrak{g}$-invariant cohomology of $M$. That is, we have isomorphisms

$$H^\bullet_D(\mathcal{C}) \cong H^\bullet(M) \qquad \text{and} \qquad H^\bullet_D(\mathcal{C}^{\mathfrak{g}}) \cong H^\bullet(M)^{\mathfrak{g}} \ . \tag{4.10}$$

The proof is based on the fact that both $\mathcal{G}$ and $W(\mathfrak{g})$ are $d$-acyclic and that the chain homotopy manifesting the acyclicity commutes with the $G$ action. This means that the $d$-cohomology of $\mathcal{C}$ (resp. $\mathcal{C}^{\mathfrak{g}}$) is precisely $H(M)$ (resp. $H(M)^{\mathfrak{g}}$). That is, there is no $d$-cohomology in ghost number different from zero. Finally, we apply either a standard spectral sequence argument or a standard descent equation argument to show that the cohomology of $D$ and $d$ agree.

An interesting fact about the double complex $(\mathcal{C}^{\bullet,\bullet},d,\delta)$ is that the $G$ action on $\mathcal{C}$ can also be written in a $\mathfrak{g}$-DGA form relative to the BRST operator. That is, there exist antiderivations $I_a$ such that $\mathcal{L}_a = [\delta, I_a]$. Indeed $I_a$ is defined uniquely by $I_a\phi = 0$ for any $\phi \in W(\mathfrak{g}) \otimes \Omega(M)$, $I_a c^b = \delta_a^b$, and $I_a dc^b = 0$. Notice that $[d, I_a] = 0$. It follows that the total complex $(\mathcal{C}^\bullet, D)$ has a one-parameter family of $\mathfrak{g}$-DGA structures with the same differential and the same $\mathcal{L}_a$. In fact, if $t$ is any parameter, $\iota_a \equiv t\imath_a + (1-t)I_a$ obeys $\mathcal{L}_a = [D, \iota_a]$ and $[\mathcal{L}_a, \iota_b] = f_{ab}{}^c \iota_c$. This means that $\mathcal{C}^\bullet$ has a one-parameter family of basic subcomplexes with a corresponding family of equivariant cohomologies. This may be interesting to pursue further.

## The Obstruction in the Noether Procedure

The Noether procedure starts with a closed form $\omega \in \Omega(M)^{\mathfrak{g}}$ and asks whether there exists an invariant closed form $\Theta \in (W(\mathfrak{g}) \otimes \Omega(M))^{\mathfrak{g}}$ such that $\delta\omega = \delta\Theta$ and such that $\Theta = 0$ when we put $A{=}F{=}0$. This suggests that the obstruction class is the image of $\omega$ under the connecting homomorphism induced by $\delta$ in a suitable complex. We now build one such complex.

Let us first introduce some notational abbreviations. We shall be dealing with some subcomplexes of $\mathcal{C}^{\mathfrak{g}}$. First we introduce the subcomplex $\Omega \equiv (\Omega(M) \otimes \mathcal{G})^{\mathfrak{g}}$. Notice that $\Omega$ inherits the bigrading from $\mathcal{C}^{p,q}$. We shall let $Z(\Omega)$ denote the $d$-cocycles and $B(\Omega)$ the $d$-coboundaries and notice that the $d$-acyclicity of $\mathcal{G}$ for positive ghost number implies that $Z^p(\Omega^{\bullet,q\geq 1}) = B^p(\Omega^{\bullet,q\geq 1})$ for all $p$. The WZ term $\omega$ of a $d$-dimensional $\sigma$-model belongs to $Z^{d+1}(\Omega^{\bullet,0})$. Similarly let $Z(\mathcal{C}^{\mathfrak{g}})$ denote the $d$-cocycles in $\mathcal{C}^{\mathfrak{g}}$. Again, the $d$-acyclicity of $\mathcal{G}$ in positive ghost number implies that $Z^p((\mathcal{C}^{\mathfrak{g}})^{\bullet,q\geq 1}) = B^p((\mathcal{C}^{\mathfrak{g}})^{\bullet,q\geq 1})$ for all $p$. Because $\delta$ anticommutes with $d$, $Z(\Omega)$, $Z(\mathcal{C}^{\mathfrak{g}})$, $B(\Omega)$, and $B(\mathcal{C}^{\mathfrak{g}})$ are closed under $\delta$ and can be considered as BRST complexes in their own right.

Now let $\Pi : \mathcal{C} \to \Omega$ denote the map corresponding to setting $A{=}F{=}0$. Since $\Pi$ commutes with $d$ it induces a map also called $\Pi$:

$$\Pi : Z^p((\mathcal{C}^{\mathfrak{g}})^{\bullet,q}) \longrightarrow Z^p(\Omega^{\bullet,q}) \; . \tag{4.11}$$

Contrary to the map $\Phi$ in (4.2), $\Pi$ is not a morphism of (BRST) complexes. This is because the BRST variation of $A$ has a term which is independent of $A$ and hence will not be annihilated by $\Pi$. Nevertheless, it is easy to verify that in any $d$-cocycle in $\mathcal{C}$, the monomial of the form $A^a\phi_a$, where $\phi_a \in \Omega(M) \otimes \mathcal{G}$, is such that $\phi_a$ is $d$-closed—in fact, $d$-exact. This means that $\Pi\delta(A^a\phi_a) = -d(c^a\phi_a)$, hence it is exact. In other words, the map $\overline{\Pi}$ defined by the following commuting diagram

$$
\begin{array}{ccccc}
 & & 0 & & \\
 & & \downarrow & & \\
 & & B^p(\Omega^{\bullet,q}) & & \\
 & & \downarrow & & \\
Z^p((\mathcal{C}^{\mathfrak{g}})^{\bullet,q}) & \xrightarrow{\Pi} & Z^p(\Omega^{\bullet,q}) & \longrightarrow & 0 \\
 & \overline{\Pi} \searrow & \downarrow & & \\
 & & H^p_d(\Omega^{\bullet,q}) & & \\
 & & \downarrow & & \\
 & & 0 & &
\end{array}
\tag{4.12}
$$

does commute with $\delta$. We shall let $\mathcal{K}$ denote its kernel. More concretely, $\mathcal{K}^{p,q}$ are those invariant $d$-cocycles $k \in \mathcal{C}^{\mathfrak{g}}$ of degree $p$ and ghost number $q$ such that $\Pi(k)$ is $d$-exact. The $d$-acyclicity of $\mathcal{G}$ imply that for $q \geq 1$, $H^p_d(\Omega^{\bullet,q}) = 0$, hence $\mathcal{K}^{p,q}$ agrees with $Z^p(\mathcal{C}^{\mathfrak{g}})^{\bullet,q}$. This means that the long exact sequence in cohomology induced by the short exact sequence of complexes:

$$
0 \longrightarrow \mathcal{K}^{p,q} \longrightarrow Z^p((\mathcal{C}^{\mathfrak{g}})^{\bullet,q}) \xrightarrow{\overline{\Pi}} H^p_d(\Omega^{\bullet,q}) \longrightarrow 0 \tag{4.13}
$$

breaks off into the five-term exact sequence below (and a series of isomorphisms which need not concern us here):

$$
\begin{array}{l}
0 \longrightarrow H^0_\delta\, \mathcal{K}^{p,\bullet} \longrightarrow H^0_\delta\, Z^p(\mathcal{C}^{\mathfrak{g}}) \xrightarrow{\overline{\Pi}_*} H^p(M)^{\mathfrak{g}} \\
\qquad \xrightarrow{\delta_*} H^1_\delta\, \mathcal{K}^{p,\bullet} \longrightarrow H^1_\delta\, Z^p(\mathcal{C}^{\mathfrak{g}}) \longrightarrow 0\ ,
\end{array}
\tag{4.14}
$$

where we have used that $H^0_\delta\, H^p(M)^{\mathfrak{g}} = H^p(M)^{\mathfrak{g}}$. Notice, moreover, that $H^0_\delta\, Z^p(\mathcal{C}^{\mathfrak{g}})$ is precisely the space of $d$-closed equivariant forms on $W(\mathfrak{g}) \otimes \Omega(M)$; that is, the closed forms in $\Omega_{\mathfrak{g}}(M)$.

Now let $\omega$ be a closed invariant $(d+1)$-form on $M$. By the above remarks, some form in the same invariant cohomology class admits a closed equivariant extension precisely when it lies in the image of $\overline{\Pi}_*$. By the exactness of (4.14), this happens precisely when it is annihilated by $\delta_*$. By the definition of the connecting homomorphism, $\delta_*[\omega] = [\delta\omega] \in H^1_\delta\, \mathcal{K}^{d+1,\bullet}$. Therefore $\delta_*[w] = 0$ precisely when there exists a $\Theta' \in \mathcal{K}^{d+1,0}$ such that $\delta\omega = \delta\Theta'$. Now, by definition of $\mathcal{K}$, $\Pi(\Theta') = d\theta$ is exact with $\theta$ invariant. Hence we find that $\delta(\omega - d\theta) = \delta(\Theta' - d\theta)$, where $\Theta \equiv \Theta' - d\theta$ is annihilated by $\Phi$. So that $\omega - d\theta$ can be gauged via the Noether method. But since $\theta$ is invariant, the $d\theta$ term can be gauged trivially by minimal coupling, hence $\omega$ can also be gauged. Conversely, if $\omega$ can be gauged via the Noether procedure, then clearly $\delta_*[\omega] = 0$. Thus we have proven the following

THEOREM 4.15. *The obstruction to gauging the WZ term $\omega \in \Omega^{d+1}(M)^{\mathfrak{g}}$ is the class $\delta_*[\omega] \in H^1_\delta\, \mathcal{K}^{d+1,\bullet}$.*

Relation with Consistent Anomalies

It is interesting to compare, if only briefly, the connection between this result and the homological treatment of consistent anomalies of **[15]**. This section is somewhat disconnected to the rest of the paper and there will be some notational conflict. We will let $E \to S$ be a principal $G$ bundle where $S$ it to be thought of as our spacetime. Let $\operatorname{Ad} E \equiv E \times_G G$ and $\operatorname{ad} E \equiv E \times_G \mathfrak{g}$ denote respectively the associated adjoint group and algebra bundles. Let $\Omega(S; \operatorname{ad} E) \equiv \Gamma(\bigwedge T^*S \otimes \operatorname{ad} E)$ denote the space of differential forms on $S$ with values in $\operatorname{ad} E$. Let $M \to S$ be the associated bundle corresponding to the matter fields of the theory perhaps twisted by some spinor bundle. We shall let $\psi \in \Gamma(M)$ denote collectively the matter fields of the theory and we let $A$ denote the gauge field associated to a connection on $E$ and we let $\mathcal{A}$ denote the space of all gauge fields. This is an affine space on which the gauge group $\mathcal{G} \equiv \Gamma(\operatorname{Ad} E)$ acts via affine transformations. Let $\operatorname{Lie} \mathcal{G} \equiv \Gamma(\operatorname{ad} E)$ denote the Lie algebra of the gauge group. The gauge group and its Lie algebra also act naturally on the space $\Gamma(M)$ of matter fields.

Now suppose that we have a theory described by a gauge-invariant action $I[\psi, A]$. Then the effective action $W[A]$ defined by

$$e^{-W[A]} \equiv Z[A] = \int \mathcal{D}\psi\, e^{-I[\psi,A]}$$

defines a function on $\mathcal{A}$. We shall refer to the space of such functions on $\mathcal{A}$ loosely as $\mathcal{F}(\mathcal{A})$. The action of the gauge group on $\mathcal{A}$ induces an action on $\mathcal{F}(\mathcal{A})$. A particularly relevant subspace of $\mathcal{F}(\mathcal{A})$ is the space of *local* functions $\mathcal{F}_{\text{loc}}(\mathcal{A})$ consisiting of those functions which are integrals over $S$ of differential polynomials of the fields. Clearly this subspace is a closed under the action of the gauge group. It is also clear that the effective action is highly nonlocal, hence it does not belong to $\mathcal{F}_{\text{loc}}(\mathcal{A})$.

The anomaly measures the failure of the effective action to be gauge-invariant; that is, to define a bona fide function on the space $\mathcal{A}/\mathcal{G}$. Indeed, if we now perform an infinitesimal gauge transformation with parameter $\xi \in \mathrm{Lie}\,\mathcal{G}$ on the effective action we obtain (assuming for simplicity that the anomalies are matter-independent)

$$\delta_\xi W[A] = \int_S \xi a(A) \ , \tag{4.16}$$

where $a(A)$ is a differential polynomial in the gauge fields. The map $\mathrm{Lie}\,\mathcal{G} \to \mathcal{F}_{\mathrm{loc}}(\mathcal{A})$ defined by $\xi \mapsto \delta_\xi W[A]$ defines a 1-cochain of the Lie algebra of the group of gauge transformations with coefficients in the module $\mathcal{F}_{\mathrm{loc}}(\mathcal{A})$, whose differential is precisely the BRST operator $\delta$. This cochain is clearly a cocycle—this being the Wess–Zumino consistency condition—since it is obtained from the gauge variation of a (nonlocal) function, and the anomaly is its cohomology class in $H^1(\mathrm{Lie}\,\mathcal{G};\mathcal{F}_{\mathrm{loc}}(\mathcal{A}))$; that is, the anomaly is inessential when it can be obtained as the gauge variation of a local "counterterm" in $\mathcal{F}_{\mathrm{loc}}(\mathcal{A})$.

We can visualize this class as follows. We consider the following short exact sequence of $\mathrm{Lie}\,\mathcal{G}$ modules:

$$0 \longrightarrow \mathcal{F}_{\mathrm{loc}}(\mathcal{A}) \longrightarrow \mathcal{F}(\mathcal{A}) \longrightarrow \overline{\mathcal{F}(\mathcal{A})} \longrightarrow 0 \ , \tag{4.17}$$

where $\overline{\mathcal{F}(\mathcal{A})} \equiv \mathcal{F}(\mathcal{A})/\mathcal{F}_{\mathrm{loc}}(\mathcal{A})$ is the space of equivalence classes of functions on $\mathcal{A}$, where two functions are said to be equivalent if their difference is a local function. The above short exact sequence yield the usual long exact sequence in cohomology:

$$\begin{array}{l} 0 \longrightarrow H^0(\mathrm{Lie}\,\mathcal{G};\mathcal{F}_{\mathrm{loc}}(\mathcal{A})) \longrightarrow H^0(\mathrm{Lie}\,\mathcal{G};\mathcal{F}(\mathcal{A})) \longrightarrow H^0(\mathrm{Lie}\,\mathcal{G};\overline{\mathcal{F}(\mathcal{A})}) \xrightarrow{\delta_*} \\ \longrightarrow H^1(\mathrm{Lie}\,\mathcal{G};\mathcal{F}_{\mathrm{loc}}(\mathcal{A})) \longrightarrow \cdots \end{array} \ . \tag{4.18}$$

The effective action $W[A]$ defines a zero-cochain $\overline{W}[A]$ in $\overline{\mathcal{F}(\mathcal{A})}$ and (4.16) says that this cochain is a cocycle. The anomaly can then be understood as the class $\delta_*[\overline{W}] \in H^1(\mathrm{Lie}\,\mathcal{G};\mathcal{F}_{\mathrm{loc}}(\mathcal{A}))$, where $\delta_*$ is the connecting homomorphism.

This result is to be compared with Theorem 4.15. Apart from the formal similarities, the relation between anomalies and the obstructions to gauging a WZ term should be well-known. In fact, it follows from the work of Hull and Spence **[6]** that in the case of an even-dimensional WZW model with target space a reductive Lie group, the obstruction to gauging the WZ term defined by the trace of an odd power of the Maurer–Cartan form is precisely the consistent chiral gauge anomaly. It would be very interesting to elucidate whether the obstruction class for a generic $\sigma$-model with WZ term can be understood as a generalized anomaly in a quantum field theory described effectively by the $\sigma$-model.

In this section we make contact with the work of Hull and Spence [**6**]. We first introduce the Cartan model for equivariant cohomology. We then expand the obstruction in Theorem 4.5 and recover the series of obstructions found by Hull and Spence in [**6**]. If the obstructions are overcome, the ambiguity in their solution can also be measured homologically and this allows us to recover and generalize the topological terms found by Hull, Roček, and de Wit in [**7**].

## Minimal Coupling and the Cartan Model

Since $\Omega_\Phi(M)$ is a subcomplex of the equivariant forms, we can gain some insight into its cohomology by studying the equivariant cohomology itself. As it is defined, $H_\mathfrak{g}(M)$ is not easy to compute. One can however fare somewhat better by using a different model for $H_\mathfrak{g}(M)$. The model in question goes by the name of the Cartan model and we will see in what follows that it is intimately linked with minimal coupling. The essential observation is that in a local gauge-invariant expression, the dependence on the connection one-forms $A^a$ is always through the covariant derivative. The $A^a$ therefore provide no real information and we should be able to do away with them. In doing so we will find that $H_\mathfrak{g}(M)$ can be computed from the invariant forms in $\mathfrak{S}\mathfrak{g}^* \otimes \Omega(M)$ with a differential which is a twisted version of the de Rham differential on $M$. We review this now following [**16**], but see also [**17**].

First of all consider the subalgebra $\bigwedge\mathfrak{g}^* \otimes \Omega(M) \subset \mathcal{C}$. Let us define a family of operations $p_a = A^a \imath_a$ (where we disable momentarily the summation convention). Here $A^a$ is the operation of multiplying by $A^a$ and $\imath_a$ is the contraction. It is easy to see that for $a \neq b$, $p_a p_b = p_b p_a$ and that $p_a^2 = p_a$. In other words, the $p_a$ are mutually commuting projectors. Hence so are their complements $1 - p_a$, and their product

$$\mathcal{P} = \prod_a (1 - p_a) \tag{5.1}$$

is a projector as well. This projector allows us to define a map

$$\begin{aligned} \kappa : \Omega(M) &\longrightarrow \bigwedge\mathfrak{g}^* \otimes \Omega(M) \\ \omega &\mapsto \mathcal{P}\,\omega\ , \end{aligned} \tag{5.2}$$

which agrees with minimal coupling. In fact, if $x^i$ are local coordinates on $M$ and $dx^i$ are the corresponding one-forms, then $\mathcal{P}\,dx^i = \nabla x^i$ given in (2.6).

Let us now define a partial inverse to minimal coupling. Let

$$\varepsilon : \bigwedge \mathfrak{g}^* \otimes \Omega(M) \to \Omega(M) \tag{5.3}$$

correspond to the operation of putting $A^a = 0$. Clearly $\varepsilon \circ \kappa$ is the identity on $\Omega(M)$, but $\kappa \circ \varepsilon$ is the projector on horizontal forms in $\bigwedge \mathfrak{g}^* \otimes \Omega(M)$. Indeed,

$$\imath_a \mathcal{P} = \prod_{b \neq a} (1 - A^b \imath_b) \cdot \imath_a (1 - A^a \imath_a) = \prod_{b \neq a} (1 - A^b \imath_b) \cdot (\imath_a - \imath_a) = 0 \ ;$$

so that the image of $\mathcal{P}$ is contained in the horizontal subalgebra. Conversely, if $\alpha$ is horizontal, then

$$\mathcal{P}\, \alpha = \prod_a (1 - A^a \imath_a) \alpha = \alpha \ ;$$

so that $\alpha$ is in the image of $\mathcal{P}$. In other words, we have proven

Lemma 5.4. *The maps $\varepsilon$ and $\kappa$ induce an algebra isomorphism*

$$\Omega(M) \cong \{\bigwedge \mathfrak{g}^* \otimes \Omega(M)\}_{\text{horizontal}} \ .$$

We can now tensor with $\mathfrak{S}\mathfrak{g}^*$ and extend the maps $\kappa$ and $\varepsilon$ in such a way that we have an isomorphism

$$\mathfrak{S}\mathfrak{g}^* \otimes \Omega(M) \cong \{W(\mathfrak{g}) \otimes \Omega(M)\}_{\text{horizontal}} \ , \tag{5.5}$$

where we have used that the $F^a$ are horizontal.

The isomorphism in Lemma 5.4 and hence the one in (5.5) are in fact isomorphisms of $\mathfrak{g}$-modules; after all $\mathcal{P}$ (and hence $\kappa$) clearly commute with the action of $\mathfrak{g}$ and so does $\varepsilon$ which entails putting $A^a = 0$. Therefore passing to $\mathfrak{g}$-invariants, we have an algebra isomorphism:

$$(\mathfrak{S}\mathfrak{g}^* \otimes \Omega(M))^{\mathfrak{g}} \cong \{W(\mathfrak{g}) \otimes \Omega(M)\}_{\text{basic}} = \Omega_{\mathfrak{g}}(M) \ . \tag{5.6}$$

We can make this into an isomorphism of differential graded algebras, by pulling back the equivariant differential from $\Omega_{\mathfrak{g}}(M)$ to $(\mathfrak{S}\mathfrak{g}^* \otimes \Omega(M))^{\mathfrak{g}}$. In other words we define a differential $d_F$ on $(\mathfrak{S}\mathfrak{g}^* \otimes \Omega(M))^{\mathfrak{g}}$ by $d_F \equiv \varepsilon \circ d \circ \kappa$. Explicitly, since $\kappa(F^a) = F^a$, and $\varepsilon$ simply puts $A^a = 0$,

$$d_F F^a = \varepsilon(dF^a) \overset{(3.8)}{=} 0 \ , \tag{5.7}$$

and for $\omega \in \Omega(M)$

$$\begin{aligned} d_F \omega &= \varepsilon(d\, \mathcal{P}\, \omega) \\ &= \varepsilon\, d\, (\omega - A^a \imath_a \omega + \cdots) \\ &= \varepsilon\, (d\omega - F^a \imath_a \omega + \cdots) \\ &= \ (d - F^a \imath_a)\, \omega \ ; \end{aligned} \tag{5.8}$$

where we have again resumed the summation convention.

The differential graded algebra $(\mathfrak{S}\mathfrak{g}^* \otimes \Omega(M))^{\mathfrak{g}}$ with the above differential $d_F$ is called the *Cartan model* for equivariant cohomology.

Recovering the Obstructions of Hull and Spence

The Cartan model for $H_{\mathfrak{g}}(M)$ restricts nicely to a model for $H_\Phi(M)$. Indeed, the maps $\varepsilon$ and $\kappa$ in (5.6) restrict to an isomorphism

$$\left(\mathfrak{S}'\mathfrak{g}^* \otimes \Omega(M)\right)^{\mathfrak{g}} \cong \Omega_\Phi(M) \ , \tag{5.9}$$

where $\mathfrak{S}'\mathfrak{g}^*$ consists of those polynomial functions on $\mathfrak{g}$ which vanish at the origin; in other words,

$$\mathfrak{S}'\mathfrak{g}^* \equiv \bigoplus_{p\geq 1} \mathfrak{S}^p\mathfrak{g}^* \ . \tag{5.10}$$

The restriction makes sense because the action of $\mathfrak{g}$ stabilizes each $\mathfrak{S}^p\mathfrak{g}^* \otimes \Omega^q(M)$ separately, and moreover it follows from the definition of the differential $d_F$ that $(\mathfrak{S}'\mathfrak{g}^* \otimes \Omega(M))^{\mathfrak{g}}$ is a subcomplex. We call it the Cartan model for $H_\Phi(M)$.

We would now like to use this to expand the obstruction defined by $d_*[\omega] \in H^{d+2}_\Phi(M)$ and thus recover the sequence of obstructions found by Hull and Spence in **[6]**. Recall that $d_*[\omega]$ is represented by the cocycle $d\widetilde{\omega} \in \Omega^{d+2}_\Phi(M)$, where $\widetilde{\omega} = \kappa(\omega)$ is the form obtained by minimally coupling $\omega$. Under the isomorphism (5.9), the cocycle $d\widetilde{\omega} = d\kappa(\omega)$ gets mapped to $\epsilon(d\kappa(\omega)) = d_F\omega$. Thus the obstruction will vanish if and only if

$$d_F\omega = d_F\Delta \qquad \text{for some} \quad \Delta \in \left(\mathfrak{S}'\mathfrak{g}^* \otimes \Omega(M)\right)^{\mathfrak{g}} \ . \tag{5.11}$$

We now analyze this equation.

Because $\omega$ is closed, $d_F\omega = -F^a\imath_a\omega$. Since $\Delta$ has degree $d{+}1$ it breaks up according to the direct sum decomposition:

$$\Delta \in \bigoplus_{\substack{p>0,q\geq 0\\ 2p+q=d+1}} (\mathfrak{S}^p\mathfrak{g}^* \otimes \Omega^q(M))^{\mathfrak{g}} \ ; \tag{5.12}$$

in other words,

$$\begin{aligned} \Delta &= F^a\Delta_a + \tfrac{1}{2}F^aF^b\Delta_{ab} + \cdots \\ &= \sum_{i=1}^{p} \tfrac{1}{i!}F^{a_1}\cdots F^{a_i}\Delta_{a_1\cdots a_i} \ , \end{aligned} \tag{5.13}$$

where the forms $\Delta_{a_1\cdots a_p} \in \Omega^{d+1-2p}(M)$ are totally symmetric in their indices. Moreover the invariance condition of the restricted Cartan model $\mathfrak{L}_a\Delta = 0$ is translated into the conditions

$$\mathfrak{L}_a \Delta_{b_1\cdots b_p} = \sum_{i=1}^{p} f_{ab_i}{}^c \Delta_{b_1\cdots b_{i-1} c b_{i+1}\cdots b_p} \ . \tag{5.14}$$

Expanding the condition $d_F\Delta = -F^a \imath_a \omega$ gives the following set of conditions:

$$-\imath_a\omega = d\Delta_a \tag{5.15}$$

$$\imath_a\Delta_b + \imath_b\Delta_a = d\Delta_{ab} \tag{5.16}$$

$$\vdots$$

$$\sum_{i=1}^{p} \imath_{a_i}\Delta_{a_1\cdots\widehat{a_i}\cdots a_p} = d\Delta_{a_1\cdots a_p} \ , \tag{5.17}$$

$$\vdots$$

where the $\widehat{\ }$ over an index denotes its omission.

We can understand these obstructions as follows. Because of the invariance of $\omega$, $\imath_a\omega$ is closed. Equation (5.15) says that this is exact. We can understand the assignment $X_a \mapsto \imath_a\omega$ as defining a map $\mathfrak{g} \to H^d(M)$. This map is easily seen to annihilate the first derived ideal $\mathfrak{g}' = [\mathfrak{g}, \mathfrak{g}]$ since because $\omega$ is closed, $f_{ab}{}^c \imath_c\omega = [\mathfrak{L}_a, \imath_b]\omega = d\imath_a\imath_b\omega$ is exact. Hence it defines a map $\mathfrak{g}/\mathfrak{g}' \to H^d(M)$ or equivalently a class in $H^1(\mathfrak{g}) \otimes H^d(M)$. If this class is trivial so that $\Delta_a$ exists obeying (5.15), then the invariance condition (5.14) shows that $\imath_a\Delta_b + \imath_b\Delta_a$ is closed so that it defines a class in $H^{d-1}(M)$. Then (5.16) says that this class is trivial. The same is true for the remaining obstructions. In fact, each $\sum_{i=1}^{p} \imath_{a_i}\Delta_{a_1\cdots\widehat{a_i}\cdots a_p}$ is closed and defines a class in $H^{d-2p}(M)$ which has to be trivial. Thus the conditions of the type (5.17) define a sequence of classes in

$$\mathfrak{S}^p\mathfrak{g}^* \otimes H^{d-2p}(M) \tag{5.18}$$

where the $p$th class is defined whenever the preceding classes are trivial.

The invariance conditions (5.14) can also be understood in terms of Lie algebra cohomology. See **[6]** and **[10]** for the details. We should emphasize however, that the obstructions are not purely topological nor purely Lie-algebraic. Equivariant cohomology is a more subtle theory in which the precise nature of the action of the algebra on the manifold matters.

New Topological Terms

We conclude this section with a brief discussion of new topological terms derivable from equivariant classes. Supposing that the obstruction is overcome and that a closed equivariant extension of $\omega$ exists, one can ask about the uniqueness of such an extension. In other words, how unique is $\widetilde{\Delta} \in \Omega_\Phi^{d+1}(M)$ such that $d\widetilde{\omega} = d\widetilde{\Delta}$. Clearly any other $\widetilde{\Delta}' \in \Omega_\Phi^{d+1}(M)$ will do equally well provided that $\widetilde{\Delta} - \widetilde{\Delta}'$ is a cocycle. In other words, we can always add to the equivariant extension $\omega^\#$ any cocycle in $\Omega_\Phi^{d+1}(M)$. If the cocycle is also a coboundary, then the change in the gauged WZ action is simply a local gauge-invariant term which is not the minimally coupled version of anything, since when we put the gauge fields to zero it vanishes. On the other hand when the cocycle is not a coboundary it defines a new topological gauge-invariant term with the property that it vanishes when the gauge fields are put to zero. Topological terms of this kind were first discovered by Hull, Roček and de Wit in **[7]**, where they constructed new topological terms for even-dimensional $\sigma$-models for nonsemisimple Lie algebras. Here we see that topological terms of this kind (which are not simply gauging of topological terms already present in the ungauged theory) are in one-to-one correspondence with $H_\Phi^{d+1}(M)$. In particular they are not restricted to even-dimensional spacetimes nor to nonsemisimple Lie algebras. Indeed, as we will see in Section 5, there are obstructions to gauging a two-dimensional WZW model based on any Lie group admitting a bi-invariant metric. This obstruction is a nontrivial class in $H_\Phi^4(M)$ which can be used to build a new topological term for a three-dimensional $\sigma$-model.

In summary, we can interpret the somewhat exotic cohomology theory $H_\Phi(M)$ physically as follows: $H_\Phi^n(M)$ is in one-to-one correspondence with the new gauge-invariant topological terms we can add to a $(n-1)$-dimensional $\sigma$-model; whereas those classes in $H_\Phi^n(M)$ which, by the exactness of (4.4), are trivial when thought of as classes in $H_\mathfrak{g}^n(M)$ correspond to the obstructions to gauging the Wess–Zumino term in a $(n-2)$-dimensional $\sigma$-model.

## §6 Two Examples

In this section we work out two simple examples of the obstructions to gauging the WZ term of a $\sigma$-model. The first example is that of a one-dimensional $\sigma$-model whose target is a symplectic manifold. The WZ term is then given by the pull-back of the symplectic form. The non-degeneracy of the symplectic form plays no role in our considerations. We choose it simply to make contact with the work of Atiyah and Bott **[9]**. Physically this system can be understood as a particle moving on $M$ interacting with a magnetic field. This model has been studied by Papadopoulos in **[12]** who also came across the obstructions to gauging in the guise of quantum-mechanical anomalies to the classical $G$ symmetry, but who also recognized them as obstructions to gauging the WZ term. The treatment here follows **[18]**.

The second example is the gauging of the standard WZ term in a two-dimensional WZW model with target manifold a Lie group admitting a bi-invariant metric. Although WZW models are generally defined on reductive Lie groups, it has been recently realized after a paper of Nappi and Witten **[19]** that more general WZW models based on Lie groups admitting a bi-invariant metric are possible. Their interest is due largely in part to the fact that, thanks to the existence of a Sugawara construction, they are quantum-mechanically conformally invariant to all orders in perturbation theory (see **[20]** and references therein).

We work out each example in a different way. The one-dimensional example is worked out in the traditional Noether method approach where the obstructions appear stepwise. The two-dimensional example, being more complex in that regard, is analyzed using directly the expanded obstructions discussed at the end of the previous section.

<u>One-dimensional $\sigma$-Model with Symplectic Target</u>

Let $B$ be a two-manifold with boundary $\partial B = \Sigma$. Let $M$ be a symplectic manifold with symplectic form $\omega \in \Omega^2(M)$. Let $\overline{\varphi} : B \to M$ and let $\varphi$ denote its restriction to $\Sigma$. The WZ term is given by

$$S_{\text{WZ}} = \int_B \overline{\varphi}^* \omega \ . \tag{6.1}$$

Suppose now that $G$ is a connected Lie group acting symplectically on $M$. The action $S_{\text{WZ}}$ is clearly invariant and we can try to gauge it. The Noether method gives us a way to analyze this problem in steps. At each step we must fulfill some condition, and this is how the obstructions arise in practice.

The $G$-invariance of $\omega$ means that for every $X \in \mathfrak{g}$, the 1-form $\imath(\xi_X)\omega$ is closed. The variation of $S_{\text{WZ}}$ under a gauge transformation with parameters $\lambda^a$ is easily worked out:

$$\begin{aligned} \delta_\lambda S_{\text{WZ}} &= \int_B \left(\lambda^a \overline{\varphi}^* \mathcal{L}_a \omega + d\lambda^a \overline{\varphi}^* \imath_a \omega\right) \\ &= \int_B d(\lambda^a \overline{\varphi}^* \imath_a \omega) \\ &= \int_\Sigma \lambda^a \varphi^* \imath_a \omega \ ; \end{aligned}$$

where we have used the notation $\mathcal{L}_a$ and $\imath_a$ for $\mathcal{L}(\xi_{X_a})$ and $\imath(\xi_{X_a})$ respectively.

We now introduce the gauge field $A = A^a X_a$ on $\Sigma$ whose gauge transformation is given by

$$\delta_\lambda A = d\lambda + [A, \lambda] \ . \tag{6.2}$$

We attempt to cancel $\delta_\lambda S_{\text{WZ}}$ with a term of the form

$$S_{\text{extra}} = \int_\Sigma A^a \varphi^* \phi_a \ . \tag{6.3}$$

It is clear that we must demand that

$$\imath_a \omega = d\phi_a \ . \tag{6.4}$$

Provided this holds, the variation of the full action $S_{\text{GWZ}} = S_{\text{WZ}} + S_{\text{extra}}$ is given by

$$\delta_\lambda S_{\text{GWZ}} = \int_\Sigma \left( [A, \lambda]^a \varphi^* \phi_a + A^a \lambda^b \varphi^* \mathfrak{L}_b \phi_a \right) \ . \tag{6.5}$$

This cancels if and only if

$$\mathfrak{L}_a \phi_b = f_{ab}{}^c \phi_c \ , \tag{6.6}$$

where $f_{ab}{}^c$ are the structure constants of $\mathfrak{g}$ in the chosen basis.

These conditions are well-known in symplectic geometry. They correspond to the existence of an equivariant moment map for the symplectic action. In fact, the $\phi_a$ are the hamiltonians corresponding to the symmetries generated by $\xi_{X_a}$. Provided they exist, condition (6.6) simply means that they represent $\mathfrak{g}$ under Poisson brackets, since $\mathfrak{L}_a \phi_b = \{\phi_a, \phi_b\}$. In other words these are the conditions that the Lie algebra morphism $X \mapsto \xi_X$ lifts to a morphism $\mathfrak{g} \to C^\infty(M)$.

In this case we can understand these obstructions simply in terms of Lie algebra cohomology of $\mathfrak{g}$ and the de Rham cohomology of $M$. Consider first the map $\mathfrak{g} \to H^1(M)$ given by $X \mapsto \imath(\xi_X)\omega$. This map annihilates the first derived ideal $\mathfrak{g}' = [\mathfrak{g}, \mathfrak{g}]$ because of the fact that the Lie bracket of two symplectic vectors fields is hamiltonian. In other words, the above map factors through a map $\mathfrak{g}/\mathfrak{g}' \to H^1(M)$. But such maps are in one-to-one correspondence with $H^1(\mathfrak{g}) \otimes H^1(M)$. Thus the first obstruction defines a class $o_1 \in H^1(\mathfrak{g}) \otimes H^1(M)$, defined by $o_1(X) = \imath(\xi_X)\omega$. If this class vanishes, then we have functions $\phi_X$ obeying $\imath(\xi_X)\omega = d\phi_X$. To analyze the second obstruction notice that $o_2(X,Y) \equiv \{\phi_X, \phi_Y\} - \phi_{[X,Y]}$ is locally constant so that it lies in $H^0(M)$. Moreover the Jacobi identities in $\mathfrak{g}$ and in $C^\infty(M)$ imply that it is a 2-cocycle $o_2([X,Y],Z) + \text{cyclic} = 0$. In other words, $c$ defines a class in $H^2(\mathfrak{g}) \otimes H^0(M)$. If this class vanishes, so that $o_2(X,Y) = -b([X,Y])$ for some $b : \mathfrak{g} \to H^0(M)$, then we can define $\widetilde{\phi}_X = \phi_X - b(X)$ so that $\{\widetilde{\phi}_X, \widetilde{\phi}_Y\} - \widetilde{\phi}_{[X,Y]}$. Notice that $d\widetilde{\phi}_X = d\phi_X$ so that the resulting hamiltonian vector field is the same.

We can summarize this in the following

Proposition 6.7. *The WZ term* (6.1) *can be gauged provided that the cohomology classes*

$$[o_1] \in H^1(\mathfrak{g}) \otimes H^1(M) \qquad \textit{and} \qquad [o_2] \in H^2(\mathfrak{g}) \otimes H^0(M)$$

*with representative cocycles*

$$o_1(X) = \imath(\xi_X)\omega \qquad \textit{and} \qquad o_2(X,Y) = \{\phi_X, \phi_Y\} - \phi_{[X,Y]}$$

*vanish, and where $o_2$ is defined provided that $[o_1]$ vanishes.*

In particular if $\mathfrak{g}$ is semisimple, the Whitehead lemmas guarantee that there are no obstructions. In [**8**], Weinstein interpreted (6.4) and (6.6) in terms of the $G$-invariant cohomology of $G \times M$, where the $G$-action on $G \times M$ is simply left multiplication in $G$. In [**9**], Atiyah and Bott reinterpreted (6.4) and (6.6) as the obstruction to finding an equivariant closed extension of the symplectic form. The reader may wish to compare the above results with Theorem 6.18 in [**9**].

WZW Model on a Lie Group with a Bi-invariant Metric

Let $M$ be a connected Lie group with self-dual Lie algebra $\mathfrak{m}$. That is, $\mathfrak{m}$ possesses an ad-invariant nondegenerate symmetric bilinear form $\langle - , - \rangle$. This gives rise to a bi-invariant metric on the Lie group. Let $\theta_L$ denote the left-invariant Maurer–Cartan $\mathfrak{m}$-valued form. Let $\psi : M \to M$ is the map taking a group element to its inverse, and define $\theta_R = \psi^*\theta_L$. Then $\theta_R$ is a right-invariant $\mathfrak{m}$-valued form on $M$. Fix a basis $\langle Z_i \rangle$ for $\mathfrak{m}$. Relative to this basis we can expand the Maurer–Cartan forms as follows $\theta_L = \theta_L^i Z_i$ and $\theta_R = \theta_R^i Z_i$, where $\theta_L^i$ and $\theta_R^i$ are respectively left- and right-invariant one-forms on $M$.

The following three-form defines the WZ term of the WZW model defined on $M$:

$$\omega = \tfrac{1}{6} \langle \theta_L , [\theta_L, \theta_L] \rangle = \tfrac{1}{6} \theta_L^i \theta_L^j \theta_L^k f_{ijk} \ , \tag{6.8}$$

where $f_{ijk} \equiv \langle Z_i , [Z_j, Z_k] \rangle$. It is manifestly left-invariant, whence $\psi^*\omega$ is right-invariant. But a simple calculation using the ad-invariance of the metric on $\mathfrak{m}$ shows that $\psi^*\omega = -\omega$, hence $\omega$ is bi-invariant. It follows that $d\omega = 0$.

Let $Z_i^L$ denote the unique left-invariant vector field on $M$ which agrees with $Z_i$ at the identity and let $Z_i^R = \psi_* Z_i^L$. It follows that $Z_i^R$ is the unique right-invariant vector field which takes the value $-Z_i$ at the identity. Because the metric on $M$ is bi-invariant, the isometry group of $M$ is $M \times M$. The action of isometry group then gives rise to a map from $\mathfrak{m} \times \mathfrak{m}$ to the vector fields on $M$, defined by $(Z_i, 0) \mapsto Z_i^L$ and $(0, Z_i) \mapsto Z_i^R$. Our choice of left- and right-invariant vector fields guarantees that this is a representation. We now let $G \subset M \times M$ be a subgroup of the isometry group. We can think of $\mathfrak{g}$ as a subalgebra of $\mathfrak{m} \times \mathfrak{m}$. If we fix a basis $\langle X_a \rangle$ for $\mathfrak{g}$, then we have $X_a = (\ell_a^i Z_i, r_a^i Z_i) \in \mathfrak{m} \times \mathfrak{m}$, where $\ell$ and $r$ define Lie algebra homomorphisms $\mathfrak{g} \to \mathfrak{m}$ via $\ell(X_a) = \ell_a^i Z_i$ and similarly for $r$. We shall let $\imath_a$ denote the degree $-1$ antiderivation on $\Omega(M)$ defined by contracting with the Killing vector field $\ell_a^i Z_i^L + r_a^i Z_i^R$, and we define the Lie derivative relative to this same vector field by $\mathcal{L}_a \equiv [d, \imath_a]$.

We now analyze the conditions to gauging the $G$-symmetry of $\omega$. Expanding the universal obstruction as we did in Section 4, we find the following three conditions:

$$\imath_a \omega = dv_a \qquad \exists\ v_a \in \Omega^2(M)\ , \tag{6.9}$$

$$\mathcal{L}_a v_b = f_{ab}{}^c v_c\ , \tag{6.10}$$

and

$$\imath_a v_b = -\imath_b v_a\ . \tag{6.11}$$

We tackle them one at a time. A simple calculation shows that

$$\imath(Z_i^L)\omega = -g_{ij} d\theta_L^j \qquad \text{and} \qquad \imath(Z_i^R)\omega = g_{ij} d\theta_R^j\ , \tag{6.12}$$

where $g_{ij} = \langle Z_i\ , Z_j \rangle$. This shows that (6.9) is satisfied with

$$v_a = g_{ij} r_a^i \theta_R^j - g_{ij} \ell_a^i \theta_L^j\ , \tag{6.13}$$

or if we think of $v$ as a linear map $\mathfrak{g} \to \Omega^2(M)$ so that $v_a = v(X_a)$, we can rewrite it as

$$v(X) = \langle r(X)\ , \theta_R \rangle - \langle \ell(X)\ , \theta_L \rangle \qquad \forall\ X \in \mathfrak{g}\ . \tag{6.14}$$

We can now compute the action of the Lie derivative $\mathcal{L}_X$ for $X \in \mathfrak{g}$ on $v(Y)$. Since the left-invariant vector fields generate right-translations and viceversa, we see that $\theta_L$ (resp. $\theta_R$) is annihilated by the Lie derivative with respect to the $Z_i^R$ (resp. $Z_i^L$). Using the invariance of the metric and the fact that $\ell$ and $r$ are homomorphisms, we find that

$$\mathcal{L}_X v(Y) = v([X, Y]) \qquad \forall\ X, Y \in \mathfrak{g}\ , \tag{6.15}$$

which is precisely (6.10). Finally we have to tackle (6.11). To this end let us define a symmetric function $(-,-) : \mathfrak{g} \times \mathfrak{g} \to C^\infty(M)$ by

$$(X, Y) = \tfrac{1}{2}\left(\imath_X v(Y) + \imath_Y v(X)\right)\ . \tag{6.16}$$

It is easy to show that $d(X, Y) = 0$ so that it is a locally constant function. And it then follows from this and (6.15) that it is in fact ad-invariant. In other words, $(-,-) \in (\mathfrak{S}^2\mathfrak{g}^*)^{\mathfrak{g}} \otimes H^0(M) \cong (\mathfrak{S}^2\mathfrak{g}^*)^{\mathfrak{g}}$ since $M$ is connected. Evaluating it at the identity, we find

$$(X,Y) = \langle r(X)\,, r(Y)\rangle - \langle \ell(X)\,, \ell(Y)\rangle \ . \tag{6.17}$$

In other words, it is the difference of the pull-backs to $\mathfrak{g}$ via $r$ and $\ell$ respectively of the metric on $\mathfrak{m}$; that is,

$$(-,-) \equiv r^* \langle -\,, -\rangle \ - \ \ell^* \langle -\,, -\rangle \ . \tag{6.18}$$

In conclusion, the $G \subset M \times M$ symmetry of the WZW model can be gauged if and only if the symmetric bilinear form (6.18) vanishes identically. For instance, we see that this is the case if $G \cong M$ is the diagonal subgroup, or if $G \subset M \times M$ is a chiral embedding (that is, one for which either $\ell$ or $r$ is zero) into an isotropic subalgebra. The former case gives rise to the coset construction, whereas the latter gives rise to the gauged WZW model description of the Drinfel'd–Sokolov reduction. On nonreductive Lie algebras, there are generically many more isotropic subalgebras and hence a larger class of possible gaugings. A more thorough discussion will appear in the second reference of **[20]**.

## §7 Vanishing Theorems

In this section we analyze in some more detail the cohomology theory $H_\Phi(M)$ in which the obstructions live. Since $\Omega_\Phi(M)$ is a subcomplex of $\Omega_{\mathfrak{g}}(M)$ we can gain some insight by analyzing $H_{\mathfrak{g}}(M)$ itself. Equivariant cohomology is not easy to compute in general, but using the topological interpretation as the cohomology of $M_G$ we can approximate it enough to be able to read off some a priori vanishing theorems in low dimension. Of course, the equivariant cohomology computed from the basic subcomplex of $W(\mathfrak{g}) \otimes \Omega(M)$ is guaranteed to coincide with the cohomology of $M_G$ only for compact $G$. Hence in order to use the topological approximation we will have to restrict ourselves to such groups. However since it is the algebraic complex that is really responsible for the obstructions to gauging, it is not completely satisfactory to have to impose such a condition on the topology of the group. Therefore we speculate that the vanishing theorems persist for semisimple groups which are not necessarily compact. In this case we cannot make use of the topological approximation, but if one assumes the existence of equivariant "minimal" models (in the sense of rational homotopy theory) for the de Rham cohomology of $M$, the extension does go through at least for semisimple $\mathfrak{g}$. We conclude by attempting to persuade the reader that equivariant minimal models do exist. We do this by outlining a few such equivariant minimal models that we have been able to construct by pedestrian means. This may be of independent interest.

Approximating $H_\Phi(M)$

Let $G$ be compact. In this case, the equivariant cohomology computed from the basic subalgebra $\Omega_{\mathfrak{g}}(M)$ of $W(\mathfrak{g}) \otimes \Omega(M)$ agrees with the cohomology of $M_G$. Although $M_G$ is not a manifold, it is sufficiently close to one—indeed, it has the homotopy type of a countable CW-complex—that we can use the singular cohomology version of Leray's theorem to approximate its cohomology from that of $BG$ and $M$. This will allow us to deduce some a priori vanishing theorems for low-dimensional $\sigma$-models.

Let $EG \to BG$ denote the universal fibration associated to $G$. This is characterized (up to homotopy) by the fact that it is a principal $G$-bundle with contractible total space. From the exact homotopy sequence associated to the fibration $G \to EG \to BG$ and the contractibility of $EG$, it follows that

$$\pi_q(BG) \cong \pi_{q-1}(G) \qquad \text{for } q \geq 1 \ . \tag{7.1}$$

In particular, since $G$ is assumed connected, $BG$ is simply connected. This will play an important role later on.

As we saw earlier, $M_G$ can be understood as the total space of a fiber bundle over $BG$ with typical fiber $M$—the $\sigma$-model target manifold. Neither the base not the total space of the fibration is strictly speaking a manifold, but they are close. In particular $BG$ is an inductive limit of finite-dimensional smooth manifolds; hence it is a countable CW-complex **[21]**. This allows us to use Leray's theorem (**[22]**, Theorem 15.11) in order to approximate the (singular) cohomology of $M_G$ in terms of the cohomology of the fiber $M$ and the cohomology of the base $BG$.

Because $BG$ is simply connected and assuming that $M$ has finite type—which represents no essential loss in generality—Leray's theorem guarantees the existence of a spectral sequence converging to $H^\bullet(M_G;\mathbb{R})$ whose $E_2$ term is given by

$$E_2^{p,q} = H^p(BG;\mathbb{R}) \otimes H^q(M;\mathbb{R}) \ . \tag{7.2}$$

Since $M$ is a smooth manifold, its simplicial cohomology with coefficients in $\mathbb{R}$ agrees with the de Rham cohomology. Finally, for a compact Lie group $G$, $H^p(BG;\mathbb{R})$ agrees with the $\mathfrak{g}$-equivariant cohomology of a point. This can be computed from the Cartan model for equivariant cohomology and is precisely

$$H^p(BG;\mathbb{R}) \cong H^p_{\mathfrak{g}}(\mathrm{pt}) \cong \begin{cases} \left(\mathfrak{S}^{p/2}\mathfrak{g}^*\right)^{\mathfrak{g}} & \text{for } p \text{ even;} \\ 0 & \text{for } p \text{ odd.} \end{cases} \tag{7.3}$$

In other words, it agrees with the Casimir ring of $\mathfrak{g}$, but with a doubling of the degrees of the generators.

We are actually interested in the cohomology $H_\Phi(M)$. This cohomology has the same spectral sequence as the equivariant cohomology except that in the $E_2$ term above, we restrict to $p \geq 1$. Therefore we have the following result

THEOREM 7.4. *For $G$ compact and connected and $M$ of finite type, there exists a spectral sequence $(E_r, d_r)$ converging to $H_\Phi(M)$ whose $E_2$ term is given by (for $p \geq 1$ and $q \geq 0$)*

$$E_2^{p,q} = \begin{cases} \left(\mathfrak{S}^{p/2}\mathfrak{g}^*\right)^{\mathfrak{g}} \otimes H^q(M) & \text{for } p \text{ even;} \\ 0 & \text{for } p \text{ odd.} \end{cases}$$

Notice moreover that $E_2^{p,q} = 0$ for $p$ odd and the same will be true for all the other $E_r$. In other words, since $d_r : E_r^{p,q} \to E_r^{p+r,q-r+1}$, all the odd $d_{2r+1}$ will be identically zero. The even maps are in general nontrivial. For instance, the differential $d_2 : E_2^{p,q} \to E_2^{p+2,q-1}$ agrees with the map induced by the twist in the de Rham differential in the Cartan model for $H_{\mathfrak{g}}(M)$. And as such it does depend explicitly on the details of the $\mathfrak{g}$ action on $M$.

Nevertheless, Theorem 7.4 has its uses. Vanishing propagates in a spectral sequence; hence if we find that $E_2^n = 0$ for some $n$, then the same will be true for $H_\Phi^n(M)$. Therefore analyzing the $E_2$ term in some detail will yield sufficient conditions (although generically not necessary) for the vanishing of $H_\Phi^n(M)$ and hence for the a priori vanishing of the obstructions to gauging the $\sigma$-model.

Some Vanishing Theorems for Compact $G$

For a $d$-dimensional $\sigma$-model, the WZ term is $(d+1)$-dimensional and the obstruction is $(d+2)$-dimensional. Let us enumerate the possible obstructions for the few lowest dimensions $d \leq 4$:

$$\begin{aligned}
E_2^2 &= (\mathfrak{g}^*)^{\mathfrak{g}} \otimes H^0(M) \\
E_2^3 &= (\mathfrak{g}^*)^{\mathfrak{g}} \otimes H^1(M) \\
E_2^4 &= \left((\mathfrak{g}^*)^{\mathfrak{g}} \otimes H^2(M)\right) \oplus \left(\left(\mathfrak{S}^2\mathfrak{g}^*\right)^{\mathfrak{g}} \otimes H^0(M)\right) \\
E_2^5 &= \left((\mathfrak{g}^*)^{\mathfrak{g}} \otimes H^3(M)\right) \oplus \left(\left(\mathfrak{S}^2\mathfrak{g}^*\right)^{\mathfrak{g}} \otimes H^1(M)\right) \\
E_2^6 &= \left((\mathfrak{g}^*)^{\mathfrak{g}} \otimes H^4(M)\right) \oplus \left(\left(\mathfrak{S}^2\mathfrak{g}^*\right)^{\mathfrak{g}} \otimes H^2(M)\right) \oplus \left(\left(\mathfrak{S}^3\mathfrak{g}^*\right)^{\mathfrak{g}} \otimes H^0(M)\right) \ .
\end{aligned}$$

The first two terms are in fact already equal to $H_\Phi^2(M)$ and $H_\Phi^3(M)$, since $d_2$ necessarily maps to zero.

We can obtain some interesting results by making a further mild assumption on $G$. A compact connected Lie group $G$ is reductive; that is, a direct product of a semisimple group and its connected center, which is a torus. If we take $G$ to be semisimple, then $[\mathfrak{g}, \mathfrak{g}] = \mathfrak{g}$ and there are no $\mathfrak{g}$-invariant elements in $\mathfrak{g}^*$. Therefore if $G$ is assumed semisimple as well, then we have that the first two cases $d = 0$ and $d = 1$ have no obstructions. We saw this already for $d = 1$ in the symplectic case where in fact, $\mathfrak{g}$ had only to satisfy that $H^1(\mathfrak{g}) = H^2(\mathfrak{g}) = 0$. The rest of the cases $d = 2$, $d = 3$ and $d = 4$ now become:

$$
\begin{aligned}
E_2^4 &= \left(\mathfrak{S}^2\mathfrak{g}^*\right)^{\mathfrak{g}} \otimes H^0(M) \\
E_2^5 &= \left(\mathfrak{S}^2\mathfrak{g}^*\right)^{\mathfrak{g}} \otimes H^1(M) \\
E_2^6 &= \left(\left(\mathfrak{S}^2\mathfrak{g}^*\right)^{\mathfrak{g}} \otimes H^2(M)\right) \oplus \left(\left(\mathfrak{S}^3\mathfrak{g}^*\right)^{\mathfrak{g}} \otimes H^0(M)\right) .
\end{aligned}
$$

From the case $d = 2$ we cannot read off anything from these results, since $E_2^4$ is definitely not zero for all $M$ and $G$: $\mathfrak{g}$ has at least one invariant symmetric bilinear form, namely the Killing form. In fact we already saw that for the case of a WZW model, there was no generic condition that would guarantee the vanishing of the obstruction to gauging the WZ term but rather we were left with the vanishing of the invariant bilinear form $(-,-) \in E_2^4$ defined by equation (6.18).

However for $d = 3$ and $d = 4$ we can read off the following results:

Corollary 7.5. *For any three-dimensional $\sigma$-model whose target manifold $M$ obeys $H^1(M) = 0$ any compact semisimple group can be gauged. This holds, in particular, if $M$ is simply connected or if $\pi_1(M)$ is perfect.*

Corollary 7.6. *For any four-dimensional $\sigma$-model whose target manifold $M$ obeys $H^2(M) = 0$, any compact semisimple group having no cubic casimir can be gauged. This is true, in particular, if the target manifold is a Lie group whose maximal compact subgroup has at most one $U(1)$ factor.*

Proof: If $\mathfrak{g}$ has no cubic casimir, then the only contribution to $E_2^6$ comes from $\left(\mathfrak{S}^2\mathfrak{g}^*\right)^{\mathfrak{g}} \otimes H^2(M)$. Since the first tensorand is not zero, this vanishes if and only if $H^2(M) = 0$. Now any Lie group has the homotopy type of its maximal compact subgroup. Over the reals, any compact group is (homologically) a product of odd spheres; hence the only contribution to $H^2(M)$ can come from $S^1 \times S^1$, where each $S^1$ comes from $U(1)$. □

It is amusing to notice that the four-dimensional $\sigma$-models which seem to play a phenomenological role **[1]**, are those with $SU(3)$ for which there is generically an obstruction, namely the chiral gauge anomaly.

The Extension to Noncompact $G$

Although the vanishing theorems obtained above used Theorem 7.4 which in turn was derived on topological grounds, the problem of gauging a Wess–Zumino term is an algebraic problem, and as such the compactness of $G$ should not play such a decisive role. In other words, it seems reasonable to expect that since only the Lie algebra $\mathfrak{g}$ enters in the definition of the equivariant complex, it is enough that the Lie algebra be semisimple—so that it has a compact real form—in order for the vanishing theorems to hold. In order to prove this, however, we must approximate $H_\Phi(M)$ algebraically. Usually topological arguments have their algebraic counterpart; but in this case we have been unable to find an algebraic spectral sequence which mimics exactly the Leray sequence. There is however an obvious algebraic spectral sequence that will be of use.

Consider the restricted Cartan model (5.9) for $H_\Phi(M)$ defined in Section 5. Let us decompose $d_F = d+\delta$, where $d$ is the de Rham differential on $\Omega(M)$ and $\delta = -F^a \imath_a$. An explicit calculation shows that $d$ and $\delta$ separately square to zero and moreover they anticommute; or in other words, that we have a double complex. We can also understand this without the need of a computation by a simple counting of degrees. Let us define the following bigraded algebra

$$K^{p,q} \equiv (\mathfrak{S}^p\mathfrak{g}^* \otimes \Omega^q(M))^{\mathfrak{g}} \ . \tag{7.7}$$

One checks that $d_F : K^{p,q} \longrightarrow K^{p,q+1} \oplus K^{p+1,q-1}$. Similarly,

$$d_F^2 : K^{p,q} \longrightarrow K^{p,q+2} \oplus K^{p+1,q} \oplus K^{p+2,q-2} \ . \tag{7.8}$$

Each piece has to vanish separately and one checks that these are respectively $d^2$, $[d,\delta]$, and $\delta^2$. If we define the total complex

$$K^n \equiv \bigoplus_{2p+q=n} K^{p,q} \tag{7.9}$$

then $d_F : K^n \to K^{n+1}$. In terms of the bigraded complex $K^{p,q}$, $d_F$ looks somewhat skewed. In order to make contact with the usual notation, it will be convenient to straighten out the bigrading by defining $J^{p,q} \equiv K^{p,q-p}$, so that $d : J^{p,q} \to J^{p,q+1}$ and $\delta : J^{p,q} \to J^{p+1,q}$. Therefore $(J^{\bullet,\bullet}, d, \delta)$ is again a double complex—albeit a more standard one. The associated total complex is of course the same as before; where now

$$K^n = \bigoplus_{\substack{p+q=n\\ q\geq p\geq 1}} J^{p,q} \ . \tag{7.10}$$

It is the cohomology of this double complex that we are interested in. Notice that even though the complex $K$ is infinite, it is bounded in each dimension;

since for a fixed $n$, $K^n$ has only a finite number of summands. Standard arguments now guarantee the existence of two spectral sequences converging to the total cohomology. We choose to analyze the spectral sequence for which $d$ is the first differential. The $E_0$ term in this spectral sequence is given simply by

$$E_0^{p,q} = J^{p,q} = \left(\mathfrak{S}^p\mathfrak{g}^* \otimes \Omega^{q-p}(M)\right)^{\mathfrak{g}} \tag{7.11}$$

with differential $d_0$ being the de Rham differential on $M$. Then the $E_1$ term is given by

$$E_1^{p,q} = H_d^{q-p}\left((\mathfrak{S}^p\mathfrak{g}^* \otimes \Omega^\bullet(M))^{\mathfrak{g}}\right) \ . \tag{7.12}$$

This involves computing the de Rham cohomology of invariant $\mathfrak{S}\mathfrak{g}^*$-valued forms. We would like to argue that under some circumstances this is the same as first computing the de Rham cohomology on $\mathfrak{S}\mathfrak{g}^*$-valued forms and *then* taking invariants. Let us proceed under this assumption and return later to a discussion of what this entails. Computing the de Rham cohomology of $\mathfrak{S}\mathfrak{g}^*$-valued forms one finds

$$H_d^{q-p}\left(\mathfrak{S}^p\mathfrak{g}^* \otimes \Omega^\bullet(M)\right) \cong \mathfrak{S}^p\mathfrak{g}^* \otimes H^{q-p}(M) \ . \tag{7.13}$$

Computing invariants and noticing that $G$—being connected—acts trivially on $H^\bullet(M)$, we find

$$\left(H_d^{q-p}\left(\mathfrak{S}^p\mathfrak{g}^* \otimes \Omega^\bullet(M)\right)\right)^{\mathfrak{g}} \cong (\mathfrak{S}^p\mathfrak{g}^*)^{\mathfrak{g}} \otimes H^{q-p}(M) \ . \tag{7.14}$$

In other words, *under the assumption that computing cohomology and taking invariants commute*, we find that the $E_1$ term of the spectral sequence is

$$E_1^{p,q} = (\mathfrak{S}^p\mathfrak{g}^*)^{\mathfrak{g}} \otimes H^{q-p}(M) \ . \tag{7.15}$$

This is (up to a regrading) the same conclusion as in Theorem 7.4, whence Corollary 7.5 and Corollary 7.6 follow without the compactness assumption.

Vanishing Theorems for Semisimple $\mathfrak{g}$ and Equivariant Minimal Models

We close this section with a discussion of the validity of inverting the order of the operations of taking cohomology and taking invariants. Let $(C, d)$ be any complex on which a Lie algebra $\mathfrak{g}$ acts in such a way that it commutes with the differential $d$. Let $Z$, $B$, and $H \equiv H(C)$ denote respectively the cocycles, coboundaries and the cohomology of $C$ relative to $d$. All three spaces are stabilized by $\mathfrak{g}$ so we can consider each one as a $\mathfrak{g}$-module in their own right. By definition of $H$ there is a short exact sequence of $\mathfrak{g}$-modules:

$$0 \longrightarrow B \longrightarrow Z \longrightarrow H \longrightarrow 0 \ . \tag{7.16}$$

By functoriality of Lie algebra cohomology, (7.16) induces a long exact sequence in cohomology:

$$\begin{array}{ccccccccc} 0 \longrightarrow & H^0(\mathfrak{g};B) & \longrightarrow & H^0(\mathfrak{g};Z) & \longrightarrow & H^0(\mathfrak{g};H) & \xrightarrow{\delta_*} & H^1(\mathfrak{g};B) & \longrightarrow \cdots \\ & \| & & \| & & \| & & & \\ & B^{\mathfrak{g}} & & Z^{\mathfrak{g}} & & H(C)^{\mathfrak{g}} & & & \end{array} \tag{7.17}$$

where $\delta_*$ is the connecting homomorphism. On the other hand, the cohomology of the $\mathfrak{g}$-invariant subcomplex of $C$ is by definition the quotient of the $\mathfrak{g}$-invariant cocycles by the $\mathfrak{g}$-invariant coboundaries; so that there is an exact sequence $0 \to B^{\mathfrak{g}} \to Z^{\mathfrak{g}} \to H(C^{\mathfrak{g}}) \to 0$, which extends (7.17) as follows:

$$\begin{array}{ccccccccc} 0 \longrightarrow & H^0(\mathfrak{g};B) & \longrightarrow & H^0(\mathfrak{g};Z) & \longrightarrow & H(C)^{\mathfrak{g}} & \xrightarrow{\delta_*} & H^1(\mathfrak{g};B) & \longrightarrow \cdots \\ & \| & & \| & & \uparrow\vdots & & & \\ 0 \longrightarrow & B^{\mathfrak{g}} & \longrightarrow & Z^{\mathfrak{g}} & \longrightarrow & H(C^{\mathfrak{g}}) & \longrightarrow & 0 & \end{array} \tag{7.18}$$

The dotted arrow is clearly well defined and injective; from where we read that $H(C^{\mathfrak{g}}) \cong H(C)^{\mathfrak{g}}$ precisely when the connecting homomorphism is the zero map; in other words, when for every invariant cohomology class we can find an invariant cocycle. A blanket assumption that guarantees this, is the vanishing of $H^1(\mathfrak{g};B)$. According to the first Whitehead lemma, this is the case if $\mathfrak{g}$ is semisimple *and* $B$ is finite-dimensional in each dimension or at least reducible into finite-dimensional submodules. Unfortunately, our $B$ is not finite-dimensional; but rather

$$B^{p,q} = \mathfrak{S}^p\mathfrak{g}^* \otimes d\Omega^{q-p-1}(M) \ . \tag{7.19}$$

Here is where the notion of minimal models comes in handy. To some extent, the de Rham complex $\Omega(M)$ is too big for what it computes. If $M$ is of finite type, then the de Rham cohomology is a finitely generated graded algebra. One should be able to compute this from a finitely generated differential graded algebra. This is precisely the idea behind the minimal models of Sullivan **[23]**. Roughly speaking, a ***minimal model*** (in the sense of rational homotopy theory) for $\Omega(M)$ is the smallest finitely and *freely* generated DGA $\mathcal{M}(M)$ which has the same cohomology algebra as the de Rham cohomology of $M$. The remarkable result of Sullivan is that the minimal model contains all the information about the rational homotopy type of the manifold $M$. Our interest in minimal models is rather more prosaic. We would like to find a $\mathfrak{g}$-equivariant minimal model which for the present purposes we define as a finitely and freely generated $\mathfrak{g}$-DGA such that its cohomology is isomorphic to the de Rham cohomology of $M$ and such that its $\mathfrak{g}$-equivariant cohomology (in the sense of Definition 3.11) is isomorphic to the $\mathfrak{g}$-equivariant cohomology of $M$. It is clear that if such equivariant minimal models exist, then $C^{p,q}$ is finite-dimensional and $H^1(\mathfrak{g}; B) = 0$, whence (7.15) is valid for semisimple $\mathfrak{g}$. Actually it follows from this last remark that what we need is not really a minimal model, but simply one of finite type.

To the best of our knowledge there does not exist any published proof of the existence of such equivariant minimal models for a Lie group—although see **[24]** for a finite group and **[25]** on related matters. We believe however that their existence should not be too difficult to establish, at least for 1-connected $M$: in fact, without any difficulty we have been able to construct "by hand" a few such models which we now briefly sketch by way of persuasion.

The first is the $S^1$-equivariant minimal model for $S^3$. The de Rham cohomology of $S^3$ is freely generated by a generator $z$ of degree 3. Since $z$ is odd, $z^2 = 0$. And in fact, $H(S^3)$ is the exterior algebra generated by $z$. This is in fact a minimal model of $S^3$—the exterior algebra of an element of degree 3 with zero differential. In contrast, for $S^2$ something more interesting happens: the de Rham cohomology of $S^2$ is not freely generated since the generator $u$ of degree 2 does not square to zero abstractly but does so in $S^2$ simply because there are no 4-forms. Hence we have a relation $u^2 = 0$ which we must "free". We do this by introducing a generator $v$ of degree 3 such that $dv = u^2$. This differential graded algebra is now freely generated and has the same cohomology as the de Rham cohomology of $S^2$. It is a minimal model for $S^2$. Parenthetically, Sullivan's theorem tells us that we can read the rational homotopy of $S^2$ from the generators of the minimal model; namely $\pi_2(S^2) \otimes \mathbb{Q} \cong \mathbb{Q}$, $\pi_3(S^2) \otimes \mathbb{Q} \cong \mathbb{Q}$, and zero everywhere else.

Let us now consider the $S^1$ action on $S^3$ defining the Hopf fibration over $S^2$. We let $\mathfrak{t}$ denote the Lie algebra of the circle group. We are interested in computing the $S^1$-equivariant cohomology of $S^3$ from a minimal model. In other words we look for a finitely and freely generated differential graded algebra $\mathcal{M} \equiv \mathcal{M}_{S^1}(S^3)$ on which we have an $S^1$-action commuting with the differential and inducing the trivial action in cohomology. Furthermore we demand that the cohomology of $\mathcal{M}$ agree with $H(S^3)$ but, more importantly, that if we are to compute the cohomology of the basic subcomplex of $W(\mathfrak{t}) \otimes \mathcal{M}$ we should recover the $S^1$-equivariant cohomology of $S^3$. Since the action is free, the equivariant cohomology is isomorphic to the cohomology of the orbit space $S^3/S^1 \cong S^2$. It is easy to convince oneself that the minimal model for $S^3$ is not $S^1$-equivariant. Indeed, let $\mathfrak{F}[z]$ denote the free graded commutative algebra generated by an element $z$ of (odd) degree 3; in this case it is an exterior algebra with one generator. A short calculation (using the Cartan model) shows that the cohomology of the basic subcomplex of $W(\mathfrak{t}) \otimes \mathfrak{F}[z]$ is isomorphic to $\mathfrak{S}\mathfrak{t}^* \otimes \mathfrak{F}[z]$ which is certainly not the de Rham cohomology of $S^2$.

Instead, let $\mathcal{M} = \mathfrak{F}[x, y, z]$ be the free differential graded algebra with generators $x$, $y$, and $z$ in degrees 1,2, and 3 respectively. In order to define an $S^1$ action on $\mathcal{M}$ we define $\mathcal{L} = [d, \imath]$, where $d$ is the differential and $\imath$ is an antiderivation of degree $-1$—the algebraic analog of contracting with the Killing vector field generating the $S^1$-action. We define them as follows: $dz = 0$, $\imath z = y$ ($\Rightarrow \imath y = 0$), $dy = 0$, $dx = y$, and $\imath x = 1$. Notice that this last relation makes sense because the action is free so that the Killing vector never vanishes. It follows from these formulas, that $\mathcal{L}$ is identically zero. The cohomology of this algebra is clearly isomorphic to $H(S^3)$. On the other hand, one can show that the cohomology of the basic subcomplex of $W(\mathfrak{t}) \otimes \mathcal{M}$ is zero in every positive degree except in degree 2 in which it is one-dimensional; that is, it is isomorphic to $H(S^2) \cong H_{S^1}(S^3)$.

Rather than sketching this proof, we sketch the proof of the analogous statement for the generalized Hopf fibrations $S^1 \to S^{2n+1} \to \mathbb{CP}_n$. Let $\mathcal{M} = \mathfrak{F}[x_1, x_2, \ldots, x_{2n+1}]$ be the free differential graded algebra with generators $x_1, x_2, \ldots, x_{2n+1}$, where the degree agrees with the subscript; and let $dx_{2n+1} = 0$, $\imath x_{2i+1} = x_{2i} = dx_{2i-1}$ for $i = 1, 2, \ldots, n$, and $\imath x_1 = 1$. Then $\mathcal{L}$ is identically zero. It is easy to see that $H(\mathcal{M})$ is freely generated by $x_{2n+1}$, and on the other hand $H_{S^1}(\mathcal{M})$ is zero except in dimensions $0, 2, \ldots, 2n$, where it is one-dimensional. The proof runs as follows. We use the Cartan model for equivariant cohomology, by which $H_{S^1}(\mathcal{M})$ can be computed from the invariants of $\mathfrak{S}\mathfrak{t}^* \otimes \mathcal{M}$ with a twisted differential, where $\mathfrak{t}$ is the Lie algebra of the circle group. If we let $t$ denote the generator of $\mathfrak{t}^*$, $\mathfrak{S}\mathfrak{t}^* \cong \mathfrak{F}[t]$, where $t$ has dimension 2. Since $S^1$ is abelian, $\mathfrak{S}\mathfrak{t}^*$ is already invariant, so that the DGA in question is $\mathfrak{F}[t, x_1, x_2, \ldots, x_{2n+1}]$ with differential $d_t = d - t\imath$. One computes $d_t x_1 = x_2 - t$, which means that we can identify $t$ and $x_2$ and drop $x_1$ from the complex. This leaves a nontrivial class $[t] \in H^2$. Now $d_t x_3 = x_4 - t^2$. Therefore we can identify $x_4$ with $t^2$ and drop $x_3$. Continuing in this fashion we identify $x_{2i}$ with $t^i$ and we drop the $x_{2i-1}$ for $i = 1, 2, \ldots, n$. We then reach $d_t x_{2n+1} = -t^{n+1}$. Hence the resulting algebra $\mathfrak{F}[t, x_{2n+1}]$ with the above differential is precisely a minimal model for $\mathbb{CP}_n$.

A further example is the fibration $H \to G \to G/H$ with $H \subset G$ compact Lie groups and $H$ acts on $G$ via left multiplication, say. If $\mathfrak{h} \subset \mathfrak{g}$ denote the respective Lie algebras, then a (generally nonminimal) $\mathfrak{h}$-equivariant model for $H(G)$ is given by the $\mathfrak{g}$-invariant subcomplex of the de Rham $\Omega(G)$; that is, by the Chevalley-Eilenberg complex $\bigwedge\mathfrak{g}^*$.

We have no doubt that for (1-connected, at least) $\mathfrak{g}$-DGAs $\mathfrak{g}$-equivariant minimal models exist; and we are confident that a proof this fact should pose no major difficulties to the experts in rational homotopy theory.

## §8 Conclusions and Outlook

Let us now recap the main points of this paper. We have seen that the topological terms of a gauged bosonic $\sigma$-model with symmetry group $G$ are in one-to-one correspondence with the $\mathfrak{g}$-equivariant ($\mathfrak{g}$ is the Lie algebra of $G$) cocycles of the target manifold. This allows us to re-interpret the obstructions found by Hull and Spence to gauging the Wess–Zumino term of a bosonic $\sigma$-model. We find that the WZ term can be gauged precisely when the closed $\mathfrak{g}$-invariant form that defines it admits an equivariant closed extension (Proposition 3.17). We then analyze this condition in two ways. One way—suggested by the Noether procedure—allows us to view the obstruction in BRST cohomology at ghost number one (Theorem 4.15), analogous to the cohomological interpretation of consistent anomalies. Another more fruitful way is to use the relation with equivariant cohomology and view the obstruction class as living in a certain subcomplex of the equivariant complex of $M$ (Theorem 4.5). If $G$ is a compact semisimple group, we can then use the topological model for equivariant cohomology to derive some vanishing theorems for low ($\leq 4$) dimensional $\sigma$-models (Corollary 7.5 and Corollary 7.6). We then argue that due to the algebraic nature of the problem, the condition on the topology of $G$ should be superfluous. We introduce a notion of an equivariant minimal model and prove that provided that the target space possesses a $\mathfrak{g}$-equivariant minimal model, the vanishing theorems persist for noncompact semisimple $G$. We then construct a few equivariant minimal models to show that the concept is not vacuous. The equivariant interpretation also allows us to recover and generalize the topological terms found by Hull, Roček, and de Wit.

At least one point in this paper deserves a deeper study, namely the geometric understanding of the complex $\Omega_\Phi(M)$ in (4.3). Its definition, although perfectly justified from our point of view, seems somewhat artificial. For the case of a compact group, since the algebraic equivariant cohomology responsible for the obstructions to gauging agrees with the topological equivariant cohomology, the cohomology classes of $M_G$ corresponding to $H_\Phi(M)$ are those whose restriction to the typical fiber is trivial. More explicitly, fix a basepoint $* \in BG$ and let $j : M \to M_G$ denote the embedding of $M$ onto the fiber over $*$. Then those classes in $H_\Phi(M)$ are precisely those classes in $H(M_G)$ which are in the kernel of $j^*$. For a noncompact $G$, we have to work with the maximal compact subgroup $K \subset G$. The classifying map $\overline{\gamma} : B \to BG$ inducing the bundles $\overline{P}$ and $\overline{P} \times_G M$ now factors through a map $\overline{\gamma}' : B \to BK$ which induces $K$-bundles on $B$. This means that the obstructions should now be analyzed in $H_K(M)$. A deeper analysis of this situation may help to refine the vanishing theorems which lead to Corollary 7.5 and Corollary 7.6.

One bonus of this understanding would be to study the obstruction to gauging (compact) Lie groups acting freely on $M$. When the $G$ action on $M$ is free, the equivariant cohomology is simply the cohomology of the quotient $M/G$. Understanding $H_\Phi(M)$ geometrically would allows us to relate it to $H(M/G)$ for free actions and this in turn would allows us to deduce vanishing theorems from the knowledge of the topology of $M/G$.

Another bonus of this analysis would be the extension of the results in this paper to finite gauge transformations. The only new ingredient in this case is that we would have to work with integral cohomology. Anticipating the quantization of the model, we would take $\omega \in H(M;\mathbb{Z})^G$ to be integral already. Then we would expect an obstruction in $H(M_G;\mathbb{Z})$ or if $G$ is not compact in $H(M_K;\mathbb{Z})$ for $K \subset G$ the maximal compact subgroup. In approximating this cohomology, we could still make use of the Leray spectral sequence for integral singular cohomology provided that $G$ were connected. If $G$ were not connected, then $BG$ would not be simply connected and the $E_2$ term would not be given by the integral version of (7.2) but rather by

$$E_2^{p,q} = H^p\left(BG; \mathcal{H}^q(M;\mathbb{Z})\right) \tag{8.1}$$

where, by definition, the presheaf $\mathcal{H}^q(M;\mathbb{Z})$ sends an open set $U \subset BG$ to $H^q(\pi^{-1}U;\mathbb{Z})$ with $\pi : M_G \to BG$. If $U$ is contractible, then $H^q(\pi^{-1}U;\mathbb{Z}) \cong H^q(M;\mathbb{Z})$; hence the presheaf is locally constant. But if $BG$ is not simply connected, it may have monodromy. In this case, we would have to analyze the action of $\pi_1(BG) \cong \pi_0(G)$ on $H^q(M;\mathbb{Z})$ to compute the $E_2^{p,q}$ term and proceed further. This seems to require a case by case analysis and it is not clear what applicable general results can be obtained.

A natural extension of the results in this paper is the study of $\sigma$-models with fermions. We will report on this topic elsewhere **[26]**.

## ACKNOWLEDGEMENTS

It is a pleasure to thank Chris Hull for drawing our attention to this topic and for many enlightening conversations. We have also benefited greatly by electronic conversations with Takashi Kimura and Jim Stasheff, who suffered through earlier versions of the results in this paper as we streamlined the presentation. We are particularly grateful to Jim Stasheff for his numerous comments on a preliminary draft of the manuscript. We would like to thank Jaap Kalkman for sending us a copy of his thesis. We are also grateful to Kristoffer Rose ⟨kris@diku.dk⟩ for help with his package XY-pic (version 2.11), which we used to typeset the diagrams in this paper. One of us (SS) would like to acknowledge the hospitality extended to her by the Physics Department at QMW and Chris Hull in particular during this past academic year. Finally, we would like to acknowledge a contract of the European Commission Human Capital and Mobility Programme which partially funded this research.